\documentclass[seceq,preprint]{ptptex}

\usepackage{graphicx}

\newcommand{\Eg}{{E_{\widetilde{g}}}}
\newcommand{\Egp}{{E_{\widetilde{g}}^\prime}}
\newcommand{\Eh}{{E_{\widetilde{h}}}}
\newcommand{\Ehp}{{E_{\widetilde{h}}^\prime}}
\newcommand{\ODR}{{\overline{D}_{\normalsize \rm R}}}
\newcommand{\OUR}{{\overline{U}_{\normalsize \rm R}}}
\newcommand{\OER}{\overline{E}_{\normalsize \rm R}}
\newcommand{\QL}{Q_{\normalsize \rm L}}
\newcommand{\DL}{{D_{\normalsize \rm L}}}
\newcommand{\UL}{{U_{\normalsize \rm L}}}
\newcommand{\UR}{{U_{\normalsize \rm R}}}
\newcommand{\DR}{D_{\normalsize \rm R}}
\newcommand{\LL}{L_{\normalsize \rm L}}
\newcommand{\EL}{E_{\normalsize \rm L}}
\newcommand{\ER}{E_{\normalsize \rm R}}

\newcommand{\PB}{p_{\normalsize \rm B}}
\newcommand{\PD}{p_{\normalsize \rm D}}
\newcommand{\Pt}{p_{\normalsize \rm \tau}}

\newcommand{\RD}{r_{\normalsize \rm D}}
\newcommand{\Rt}{r_{\normalsize \rm \tau}}
\newcommand{\KMtreedag}{{V_{\normalsize \rm CKM}^{0\dag}}}
\newcommand{\KMtree}{{V_{\normalsize \rm CKM}^0}}
\newcommand{\KMdag}{{V_{\normalsize \rm CKM}^{\dag}}}
\newcommand{\KM}{{V_{\normalsize \rm CKM}}}

\notypesetlogo                       
\preprintnumber[3cm]{
IU-MSTP/62\\KEK-TH-985\\ September, 2004}

\title{
Tauonic B decays in the Minimal Supersymmetric Standard Model%
}

\author{
Hideo \textsc{Itoh}$^1$%
, Shinji \textsc{Komine}$^2$%
and Yasuhiro \textsc{Okada}$^{2,3}$%
}

\inst{
$^1$Graduate School of Science and Engineering, Ibaraki University, Mito 310-8512, Japan
\\
$^2$Theory Group, High Energy Accelerator Research Organization (KEK), Tsukuba, Ibaraki 305-0801, Japan
\\
$^3$Department of Particle and Nuclear Physics, the Graduate University for Advanced Studies (SOKENDAI), Tsukuba, Ibaraki 305-0801, Japan
}

\abst{
We study new physics effects on B decay processes including a final $\tau$ particle, namely $B \to D \tau \nu$ and $B \to \tau \nu$. An important feature of these processes is that a charged Higgs boson can contribute to the decay amplitude at the tree level in models such as Two Higgs Doublet Model and the Minimal Supersymmetric Standard Model (MSSM). We derive a resummed effective Lagrangian for charged-Higgs mediated interactions in the MSSM with the Minimal Flavor Violation. Including supersymmetric (SUSY) loop corrections for down-type-quark and charged-lepton Yukawa couplings, we calculate the branching ratios of the $B \to D \tau \nu$ and $B \to \tau \nu$ processes. 
We find that SUSY correction due to gluino-sbottom diagrams can change the Higgs exchange contribution by $\pm50$\%, whereas stau-neutralino diagrams can make corrections up to 20\%. We also discuss relationship between SUSY corrections in the tauonic decays and flavor changing neutral current processes such as $B_s \to \mu^+\mu^-$ and $b \to s \gamma$.
}

\begin{document}

\maketitle

\newpage
\section{Introduction}

Recent success of B factory experiments at KEK and SLAC has proved that B physics provide a very promising way to explore physics in and beyond the Standard Model (SM). The Kobayashi-Maskawa mechanism \cite{Kobayashi:1973fv} of the CP violation in the quark sector has been established from the precise determination of the CP asymmetry in $B \to J/\psi K_S$ and related modes \cite{Abashian:2001pa, Aubert:2001sp}. B factory experiments have made many new observations such as the branching ratio of the $b \to sll$ \cite{Kaneko:2002mr} and CP violation in the $B \to \phi K_S$ mode \cite{Abe:2003yt, Aubert:2004ii}, which are known to be sensitive to new physics effects.

In future, more information on B decays will be obtained at current B factories as well as hadron B experiments at Tevatron and LHC. Furthermore, the future upgrade of the $e^+e^-$ asymmetric B factory, Super B Factory, is discussed, where the goal of the luminosity is 50-100 times more than the current achieved luminosity \cite{L.O.I.}.

We study here new physics effects on B decay processes including a final $\tau$ particle, namely $B \to D \tau \nu$ and $B \to \tau \nu$. An important feature of these processes is that a charged Higgs boson can contribute to the decay amplitude at the tree level in models such as Two Higgs Doublet Model (2HDM) and the Minimal Supersymmetric Standard Model (MSSM). From the experimental side,  since at least two neutrinos are present in the final state in the signal side, full-reconstruction is required for the B decay in the opposite side. For the $B \to D \tau \nu$ process, the branching ratio is not yet measured even though the SM prediction is $8 \times 10^{-3}$. (The inclusive $b \to c \tau \nu$ branching ratio was determined at LEP experiments \cite{Abbiendi:2001fi}.) The $B \to \tau \nu$ process has a smaller branching ratio ($\sim 9 \times 10^{-5}$) in the SM, because of the helicity suppression, and the upper bound of $2.9 \times 10^{-4}$ is reported from the BELLE experiment \cite{Abe2} and $3.3 \times 10^{-4}$ from the BaBar \cite{Aubert}. These processes will be important targets of coming B factory experiments.

In this paper, we calculate the branching ratio of $B \to D \tau \nu$ and $B \to \tau \nu$ processes in the MSSM, taking account of supersymmetric (SUSY) corrections to the charged Higgs Yukawa couplings. At the tree level, the Higgs sector of the MSSM is of the same form as the type I{}I 2HDM, where one Higgs doublet provides mass terms for up-type quarks, and the other does for down-type quarks and charged leptons. SUSY loop corrections, however, can induce Yukawa couplings of the opposite type \cite{Hall:1993gn, Babu:1998er, Carena:1999py}. In particular, it is known that new contributions to the bottom and tau Yukawa coupling constants induce flavor changing processes such as $B_s \to \mu^+\mu^-$ \cite{FCNC, Dedes:2002er}, $b \to sll$ \cite{Huang:1998vb}, $\tau \to 3\mu$ \cite{Babu:2002et, Dedes:2002rh}, $\tau \to \mu \eta$ \cite{Sher:2002ew} and the $\mu$-$e$ conversion at muonic atoms \cite{Kitano:2003wn}, especially for a large ratio of two vacuum expectation values ($\tan\beta$).
For the $b \to c(u)$ transition, SUSY corrections were investigated in the inclusive $b \to c \tau \nu$ \cite{Coarasa:1997uw} and the $B \to \tau \nu$ processes \cite{Akeroyd:2003zr}. In this paper, in addition to the correction to the charged Higgs boson($H^\pm$)-$b$-$c(u)$ vertex, we include the correction to the $H^\pm$-$\tau$-$\nu$ vertex, and study importance of these corrections to these two processes. As for the flavor mixing in the squark and slepton sectors, we take the assumption of the Minimal Flavor Violation (MFV), where a unique origin of the flavor mixings is given by the Cabibbo-Kobayashi-Maskawa (CKM) matrix \footnote{For general SUSY models, the tauonic B decay is considered in G.H.Wu et al \cite{Wu:1997ua}.}
. We show that the corrections to the $B \to D \tau \nu$ and $B \to \tau \nu$ branching ratios are large for $\tan\beta \gtrsim 30$. The tau vertex correction can give sizable effects for reasonable parameter sets of squark and sleptons. We also consider correlation of these processes to $b \to s \gamma$ and $B_s \to \mu^+\mu^-$.

This paper is organized as follows. In section 2, we derive effective Yukawa interactions of the charged Higgs boson  and quarks/leptons taking account of SUSY loop corrections in the MSSM. We apply this formulation to calculate the $B \to D \tau \nu$ and $B \to \tau \nu$ branching ratios. In section 3, results of numerical calculations are presented including the $b \to s \gamma$ and $B_s \to \mu^+\mu^-$. Conclusions are given in section 4. Derivation of the charged Higgs coupling in the MFV case is given in Appendix A.

\section{Formalism}

\subsection{SUSY loop corrections to Yukawa interactions}
In this section, we derive the general form of the effective Lagrangian for the charged Higgs Yukawa coupling in the MSSM taking account of SUSY loop diagrams. For the flavor mixing in the squark sector, we take a model based on the assumption of MFV, where the CKM matrix is the only source of flavor and CP violations. As we see below, there is a new contribution to the Yukawa coupling constant in a large $\tan\beta$ regime. We resum the $\tan\beta$-enhanced contributions consistently for the down-type and charged-lepton Yukawa coupling constants following the method developed in the reference by A.Dedes and A.Pilaftsis \cite{Dedes:2002er}. 

The superpotential of the model is given by

\begin{eqnarray}
W = -H_1 D^c {\bf y}_d Q + H_2 U^c {\bf y}_u Q - H_1 E^c {\bf y}_e L + \mu H_1 H_2,
\end{eqnarray}
where the components of weak doublet fields are denoted as
\begin{eqnarray}
H_1 = \left( \begin{array}{c}
				H_1^0 \\
				H_1^- \\
			 \end{array}
	  \right),\ \ \ 
H_2 = \left( \begin{array}{c}
				H_2^+ \\
				H_2^0 \\
			 \end{array}
	  \right),\ \ \ 
Q = \left( \begin{array}{c}
				U \\
				D \\
			 \end{array}
	  \right),\ \ \ 
L = \left( \begin{array}{c}
				N \\
				E \\
			 \end{array}
	  \right).
\end{eqnarray}
The quantum numbers of ${\rm SU}(3)\times{\rm SU}(2)\times{\rm U}(1)$ gauge groups for $H_1,\ H_2,\ Q,\ L,\ D^c,\ U^c,\ E^c$ are $({\bf 1}, {\bf 2}, -1)$, $({\bf 1}, {\bf 2}, 1)$, $({\bf 3}, {\bf 2}, \frac{1}{3})$, $({\bf 1}, {\bf 2}, -1)$, $({\bf 3}, {\bf 1}, \frac{2}{3})$, $({\bf 3}, {\bf 1}, -\frac{4}{3})$, ({\bf 1}, {\bf 1}, 2). The soft SUSY breaking masses and trilinear SUSY breaking terms(A-term) are given by
\begin{eqnarray}
\mathcal{L}_{\rm \normalsize soft} &=& - \widetilde{Q}_{\rm \normalsize L}^\dag M_{\widetilde{Q}_{\rm \normalsize L}}^2  \widetilde{Q}_{\rm \normalsize L} - \widetilde{U}_{\rm \normalsize R}^\dag M_{\widetilde{U}_{\rm \normalsize R}}^2 \widetilde{U}_{\rm \normalsize R} - \widetilde{D}_{\rm \normalsize R}^\dag M_{\widetilde{D}_{\rm \normalsize R}}^2 \widetilde{D}_{\rm \normalsize R} - \widetilde{L}_{\rm \normalsize L}^\dag M_{\widetilde{L}_{\rm \normalsize L}}^2 \widetilde{L}_{\rm \normalsize L} - \widetilde{E}_{\rm \normalsize R}^\dag M_{\widetilde{E}_{\rm \normalsize R}}^2 \widetilde{E}_{\rm \normalsize R} \nonumber \\
&& + H_1 \widetilde{D}_{\rm \normalsize R}^\dag {\bf A}_d \widetilde{Q}_{\rm \normalsize L} - H_2 \widetilde{U}_{\rm \normalsize R}^\dag {\bf A}_u \widetilde{Q}_{\rm \normalsize L} + H_1 \widetilde{E}_{\rm \normalsize R}^\dag {\bf A}_e \widetilde{L}_{\rm \normalsize L} + {\rm h.c.},
\end{eqnarray}
where the fields with tilde ( $\widetilde{}$ ) denote squarks and sleptons.

In this section, we first discuss the simplest case where soft breaking masses are proportional to a unit matrix in the flavor space, and ${\bf A}_u, {\bf A}_d$ and ${\bf A}_e$ are proportional to Yukawa couplings. Their explicit forms are given below,
\begin{eqnarray}
M_{\widetilde{Q}_{{\rm \normalsize L} ij}}^2 &=& a_1 \widetilde{M}^2 \left( \begin{array}{ccc}
													1 && \\
													& 1 & \\
													&& 1 \\
													\end{array}\right) \equiv \left( \begin{array}{ccc}
			M_{{\widetilde{Q}_{{\rm \normalsize L} 1}}}^2 && \\
			& M_{{\widetilde{Q}_{{\rm \normalsize L} 2}}}^2 & \\
			&& M_{{\widetilde{Q}_{{\rm \normalsize L} 3}}}^2 \\
	 \end{array} \right), \label{squark mass1} \\
M_{\widetilde{U}_{{\rm \normalsize R} ij}}^2 &=& a_2 \widetilde{M}^2 \left( \begin{array}{ccc}
													1 && \\
													& 1 & \\
													&& 1 \\
													\end{array}\right) \equiv \left( \begin{array}{ccc}
			M_{{\widetilde{U}_{{\rm \normalsize R} 1}}}^2 && \\
			& M_{{\widetilde{U}_{{\rm \normalsize R} 2}}}^2 & \\
			&& M_{{\widetilde{U}_{{\rm \normalsize R} 3}}}^2 \\
	 \end{array} \right), \\
M_{\widetilde{D}_{{\rm \normalsize R} ij}}^2 &=& a_3 \widetilde{M}^2 \left( \begin{array}{ccc}
													1 && \\
													& 1 & \\
													&& 1 \\
													\end{array}\right) \equiv \left( \begin{array}{ccc}
			M_{{\widetilde{D}_{{\rm \normalsize R} 1}}}^2 && \\
			& M_{{\widetilde{D}_{{\rm \normalsize R} 2}}}^2 & \\
			&& M_{{\widetilde{D}_{{\rm \normalsize R} 3}}}^2 \\
	 \end{array} \right), \label{squark mass2} \\
M_{\widetilde{L}_{{\rm \normalsize L} ij}}^2 &=& a_4 \widetilde{M}^2 \left( \begin{array}{ccc}
													1 && \\
													& 1 & \\
													&& 1 \\
													\end{array}\right) \equiv \left( \begin{array}{ccc}
			M_{{\widetilde{L}_{{\rm \normalsize L} 1}}}^2 && \\
			& M_{{\widetilde{L}_{{\rm \normalsize L} 2}}}^2 & \\
			&& M_{{\widetilde{L}_{{\rm \normalsize L} 3}}}^2 \\
	 \end{array} \right), \label{squark mass2} \\
M_{\widetilde{E}_{{\rm \normalsize R} ij}}^2 &=& a_5 \widetilde{M}^2 \left( \begin{array}{ccc}
													1 && \\
													& 1 & \\
													&& 1 \\
													\end{array}\right) \equiv \left( \begin{array}{ccc}
			M_{{\widetilde{E}_{{\rm \normalsize R} 1}}}^2 && \\
			& M_{{\widetilde{E}_{{\rm \normalsize R} 2}}}^2 & \\
			&& M_{{\widetilde{E}_{{\rm \normalsize R} 3}}}^2 \\
	 \end{array} \right), \label{squark mass2} \\
{\bf A}_{u ij} &=& A_u {\bf y}_{u ij}, \label{A-term1} \\
{\bf A}_{d ij} &=& A_d {\bf y}_{d ij}, \label{A-term2} \\
{\bf A}_{e ij} &=& A_e {\bf y}_{e ij},
\end{eqnarray}
where $a_i(i = 1 - 5)$ are real parameters. 

At the tree level, the Yukawa couplings have the same structure as the above superpotential, namely, $H_1$ couples to $D^c$ and $E^c$, and $H_2$ to $U^c$. On the other hand, different types of couplings are induced when we take into account SUSY breaking effects through one-loop diagrams. Lagrangian of the Yukawa sector can be written as
\begin{eqnarray}
\mathcal{L}_{\normalsize \rm Yukawa} &=& -H_1 \ODR {\bf y}_d \QL + H_2 \OUR {\bf y}_u \QL - H_1\OER {\bf y}_e \LL \nonumber \\
&& - \widetilde{H}_2 \ODR \Delta {\bf y}_d \QL + \widetilde{H}_1 \OUR \Delta {\bf y}_u \QL - \widetilde{H}_2 \OER \Delta {\bf y}_e \LL + \textrm{h.c.}. \label{lagrangian}
\end{eqnarray}
where $\widetilde{H}_{1,2} \equiv i \sigma_2 H_{1,2}^*$, and $\Delta {\bf y}_d$, $\Delta {\bf y}_u$, and $\Delta {\bf y}_e$ are one-loop induced coupling constants. (Here and in the followings, quark and lepton fields in capital letters represent three vectors in the flavor space.)

From the above Yukawa couplings, we can derive the quark and lepton mass matrices and their charged Higgs couplings. For the quark sector, we get
\begin{eqnarray}
\mathcal{L}_{\rm quark} &=& - \frac{v}{\sqrt{2}}\cos\beta\ODR {\bf y}_d[1 + \tan\beta \Delta_{m_d}]\DL + \sin\beta H^-\ODR {\bf y}_d[1 - \cot\beta\Delta_{m_d}]\UL \nonumber \\
&& - \frac{v}{\sqrt{2}}\sin\beta\OUR {\bf y}_u[1 - \cot\beta\Delta_{m_u}]\UL + \cos\beta H^+\OUR {\bf y}_u[1 + \tan\beta \Delta_{m_u}]\DL + {\rm h.c.}, \nonumber \\ \label{massterm}
\end{eqnarray}
where we define $\Delta_{m_d}(\Delta_{m_u})$ as $\Delta_{m_d} \equiv {\bf y}_d^{-1}\Delta {\bf y}_d$ $(\Delta_{m_u} \equiv {\bf y}_u^{-1}\Delta {\bf y}_u)$, and $v \simeq 246$GeV. Notice that $\Delta {\bf y}_d$ is proportional to ${\bf y}_d$ or ${\bf y}_d {\bf y}_u^\dag{\bf y}_u$ in this case. We then rotate the quark bases as follows:
\begin{eqnarray}
\UL = V_{\normalsize \rm L}(Q)\UL^\prime &,&\ \ \ \DL = V_{\normalsize \rm L}(Q)V_{\normalsize \rm CKM} \DL^\prime ,\\
U_{\normalsize \rm R} = V_{\normalsize \rm R}(U)U_{\normalsize \rm R}^\prime &,&\ \ \ D_{\normalsize \rm R} = V_{\normalsize \rm R}(D)D_{\normalsize \rm R}^\prime,
\end{eqnarray}
where the fields with a prime ( $^\prime$ ) are mass eigenstates. In this basis, the down-type quark Lagrangian is given by
\begin{eqnarray}
\mathcal{L}_{\rm D-quark} &=& - \frac{v}{\sqrt{2}}\cos\beta\ODR^\prime V_{\normalsize \rm R}^\dag(D) {\bf y}_d V_{\normalsize \rm L}(Q)\hat{R}_d V_{\normalsize \rm CKM}\DL^\prime \nonumber \\
&& + \sin\beta H^-\ODR^\prime V_{\normalsize \rm R}^\dag(D) {\bf y}_d V_{\normalsize \rm L}(Q)\UL^\prime + {\rm h.c.}, \label{eq:massterm}
\end{eqnarray}
where $\hat{R}_d \equiv 1 + \tan\beta\hat{\Delta}_{m_d}$ and $\hat{\Delta}_{m_d} \equiv V_{\rm \normalsize L}^\dag(Q) \Delta_{m_d} V_{\rm \normalsize L}(Q)$. Hereafter, a matrix with a hat ( $\hat{}$ ) represents a diagonal matrix. Since the down-type diagonal mass term is given by
\begin{eqnarray}
\hat{M}_d \equiv \frac{v}{\sqrt{2}}\cos\beta V_{\normalsize \rm R}^\dag(D) {\bf y}_d V_{\normalsize \rm L}(Q)\hat{R}_dV_{\normalsize \rm CKM},
\end{eqnarray}
we obtain the following Lagrangian for down-type quarks.
\begin{eqnarray}
\mathcal{L}_{\rm D-quark} = - \ODR^\prime\hat{M}_d\DL^\prime + \frac{\sqrt{2}}{v}\tan\beta H^- \ODR^\prime\hat{M}_d V_{\normalsize \rm CKM}^\dag \hat{R}_d^{-1}\UL^\prime + {\rm h.c.}. \label{down}
\end{eqnarray}

The corresponding corrections to the up-type couplings can be calculated from Eq.(\ref{massterm}). Since we are interested in the large $\tan\beta$ case, these corrections are very small. In the following, we neglect such corrections, and the Lagrangian for the up-type-quarks is given as follows: 
\begin{eqnarray}
{\cal L}_{\rm U-quark} = -\OUR^\prime \hat{M}_u \UL^\prime + \frac{\sqrt{2}}{v}\cot\beta H^+ \OUR^\prime \hat{M}_u V_{\rm \normalsize CKM} \DL^\prime + {\rm h.c.}. \label{up}
\end{eqnarray}

For the case of the charged-lepton, we can derive relevant parts of the Lagrangian in a similar way to the case of the down-type quark. Using the following definitions,
\begin{eqnarray}
\EL &=& V_{\rm \normalsize L}(L) \EL^\prime,\ \ \ \ER = V_{\rm \normalsize R}(E) \ER^\prime, \\
\Delta_{m_e} &\equiv& {\bf y}_e^{-1}\Delta {\bf y}_e,\ \ \ \hat{\Delta}_{m_e} \equiv V_{\rm \normalsize L}^\dag(L) \Delta_{m_e}V_{\rm \normalsize L}(L),
\end{eqnarray}
we obtain the Lagrangian for the charged lepton as follows:
\begin{eqnarray}
\mathcal{L}_{\rm lepton} = - \OER^\prime\hat{M}_e\EL^\prime + \frac{\sqrt{2}}{v}\tan\beta H^-\OER^\prime\hat{M}_e\hat{R}_e^{-1}N_{\rm \normalsize L} + {\rm h.c.}. \label{lepton}
\end{eqnarray}
Here, the prime represents the mass eigenstate, and we neglect the neutrino masses, and $\hat{R}_e \equiv 1 + \tan\beta\hat{\Delta}_{m_e}$.

\begin{figure}[t]
\hspace{1.2cm}\includegraphics[width=.4\linewidth]{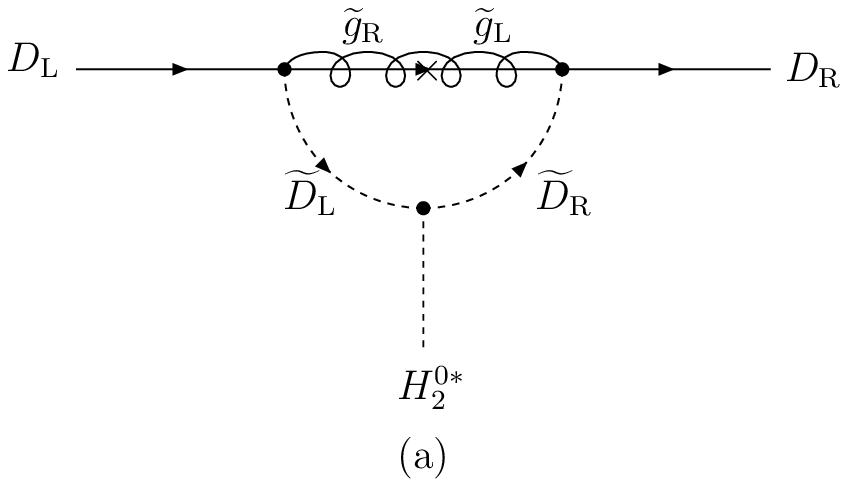}
\hspace{1.2cm}\includegraphics[width=.4\linewidth]{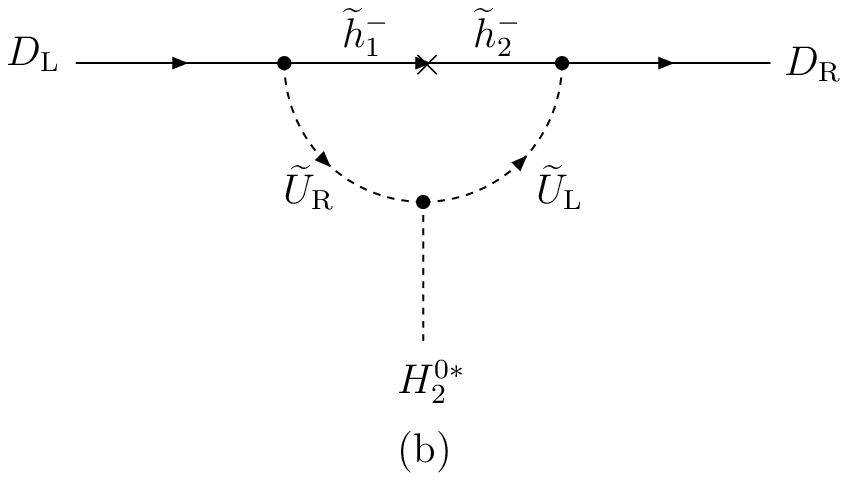}	
\caption{Non-holomorphic radiative corrections to the down-type quark Yukawa couplings induced by (a) gluino $\widetilde{g}_{\normalsize \rm L,R}$ and (b) charged higgsino $\widetilde{h}^-_{1,2}$.}\label{fig:gluino}
\hspace{5cm}\includegraphics[width=.4\linewidth]{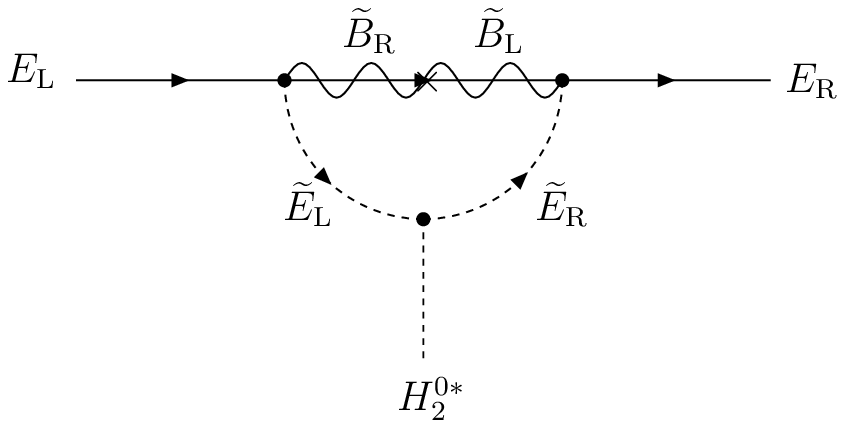}
\caption{Non-holomorphic radiative corrections to the charged lepton Yukawa couplings induced by bino $\widetilde{B}_{\normalsize \rm L,R}$.}\label{fig:bino}
\end{figure}

In the present case with Eq.(\ref{squark mass1}) - Eq.(\ref{A-term2}), $\hat{\Delta}_{m_d}$ receives contributions from gluino and down-type squark, and higgsino and up-type squark diagrams. The explicit form is given as follows:
\begin{eqnarray}
\hat{\Delta}_{m_d} = \hat{{\bf E}}_{\widetilde{g}} + \hat{{\bf E}}_{\widetilde{h}},
\end{eqnarray}
where
\begin{eqnarray}
\hat{{\bf E}}_{\widetilde{g}} &\equiv& \frac{2\alpha_s}{3\pi}{\bf 1}\mu^*M_{\tilde{g}}I[M_{\tilde{g}}, M_{\tilde{D}_{\rm \normalsize L}}, M_{\tilde{D}_{\rm \normalsize R}}], \label{gluino loop} \\
\hat{{\bf E}}_{\widetilde{h}} &\equiv& - \frac{\mu}{16\pi^2}A_{\rm \normalsize u}|\hat{{\bf y}}_{\rm \normalsize u}|^2I[M_{\widetilde{h}}, M_{\widetilde{U}_{\rm \normalsize L}}, M_{\widetilde{U}_{\rm \normalsize R}}], \label{A-term dependence} \label{higgsino loop} \\
I[a,b,c] &=& \frac{a^2b^2\ln\frac{a^2}{b^2} + b^2c^2\ln\frac{b^2}{c^2} + c^2a^2\ln\frac{c^2}{a^2}}{(a^2 - b^2)(b^2 - c^2)(a^2 - c^2)}.
\end{eqnarray}
$\hat{\bf E}_{\widetilde{g}}$ and $\hat{\bf E}_{\widetilde{h}}$ are gluino and charged higgsino contributions shown in Fig.\ref{fig:gluino}(a) and (b), respectively. Note that these corrections for Yukawa couplings are calculated in the unbroken phase of ${\rm SU}(2)\times {\rm U}(1)$ symmetry. For the charged lepton case, $\hat{\Delta}_{m_e}$ is given by

\begin{eqnarray}
\hat{\Delta}_{m_e} = \hat{{\bf E}}_{\widetilde{B}} = \frac{(M_Z^2 - M_W^2)}{4v^2\pi^2}{\bf 1}\mu M_{\widetilde{B}}I[M_{\widetilde{B}}, M_{\widetilde{L}_{\rm \normalsize L}}, M_{\widetilde{L}_{\rm \normalsize R}}], \label{bino loop}
\end{eqnarray}
from the bino-slepton diagram shown in Fig.\ref{fig:bino}.

Up to now, we have assumed all squark mass matrices are proportional to a unit matrix at the electro-weak scale, as shown in Eq.(\ref{squark mass1}) - Eq.(\ref{squark mass2}). However models with the MFV correspond to more general cases. For instance, the assumption of Eq.(\ref{squark mass1}) - Eq.(\ref{squark mass2}) is not satisfied in the minimal supergravity where all squarks have a universal mass at the Planck scale, not at the electro-weak scale. In appendix we derive the charged Higgs coupling in more general case of MFV. Namely the squark mass matrix is taken to be 
\begin{eqnarray} 
M_{{\widetilde{Q}_{\rm \normalsize L}}}^2 &=& [a_1 {\bf 1} + b_1 {\bf y}_u^\dag {\bf y}_u + b_2 {\bf y}_d^\dag {\bf y}_d]\widetilde{M}^2, \label{assumption1} \\
M_{{\widetilde{U}_{\rm \normalsize R}}}^2 &=& [a_2 {\bf 1} + b_5 {\bf y}_u {\bf y}_u^\dag]\widetilde{M}^2, \label{assumption2} \\
M_{{\widetilde{D}_{\rm \normalsize R}}}^2 &=& [a_3 {\bf 1} + b_6 {\bf y}_d {\bf y}_d^\dag]\widetilde{M}^2 \label{assumption3}.
\end{eqnarray}
The final results of the charged Higgs coupling is given by 
\begin{eqnarray}
\mathcal{L}_{H^\pm} &\approx& \frac{\sqrt{2}}{v}\tan\beta H^- \ODR^\prime_i \frac{\hat{M}_{d i}}{ 1 + [\Eg^{(i)} ] \tan\beta} \KMdag_{ij} \UL_j^\prime + {\rm h.c.} \ \ \ \ \ \ \textrm{for $(i,j)=(1,1),(1,2),(2,1),(2,2)$}, \nonumber \\
\\
\mathcal{L}_{H^\pm} &\approx& \frac{\sqrt{2}}{v}\tan\beta H^- \ODR^\prime_i \frac{\hat{M}_{d i}}{ 1 + [\Eg^{(i)} - \Egp^{(ij)} ] \tan\beta} \KMdag_{ij} \UL_j^\prime + {\rm h.c.}  \ \ \ \ \ \ \textrm{for $(i,j)=(3,1),(3,2)$}, \nonumber \\
\\
\mathcal{L}_{H^\pm} &\approx& \frac{\sqrt{2}}{v}\tan\beta H^- \ODR^\prime_i \frac{\hat{M}_{d i}}{1 + \Eg^{(i)}\tan\beta } \frac{ 1 + [ \Eg^{(3)} + \Eh^{(33)} ]\tan\beta }{1 + [ \Eg^{(i)} + \Eh^{(33)} + \Egp^{(ij)} + \Eh^{(i3)} + \Ehp^{(i33)} ]\tan\beta} \KMdag_{ij} \UL_j^\prime \nonumber \\
&&  \ \ \ \ \ \ \ \ \ \ \ \ \ \ \ \ \ \ \ \ \ \ \ \ \ \ \ \ \ \ \ \ \ \ \ \ \ \ \ \ \ \ \ \ \ \ \ \ \ \ \ \ \ \ \ \ \ \ \ \ \ \ \ \   + {\rm h.c.}\ \ \ \ \ \  \textrm{for $(i,j)=(1,3),(2,3)$}, \\
\mathcal{L}_{H^\pm} &\approx& \frac{\sqrt{2}}{v}\tan\beta H^- \ODR^\prime_i \frac{\hat{M}_{d i}}{ 1 + [\Eg^{(i)} + \Eh^{(i3)} ] \tan\beta} \KMdag_{ij} \UL_j^\prime + {\rm h.c.}\ \ \ \ \ \ \textrm{for $(i,j)=(3,3)$}.
\end{eqnarray}
where the function $\Eg^{(i)}$, etc are listed in Appendix A. In deriving these results we only keep $y_t$ in the up-type Yukawa coupling in loop diagrams and use the hierarchy of the CKM matrix elements. See Appendix A for details. Notice that the above results do not depend on the relation between the A-terms and the Yukawa couplings, since we only keep $y_t$ in loop diagrams, even though Eqs.(\ref{A-term1}) and (\ref{A-term2}) are assumed in Appendix A.

\subsection{Effective Lagrangian for 4-Fermi interactions}
Once we obtain the charged Higgs and fermion couplings, it is straightforward to write down the amplitudes for $B \to D \tau \nu\ (B^- \to \overline{D}^0 \tau^- \overline{\nu}\ {\rm or}\ \overline{B}^0 \to D^+ \tau^- \overline{\nu})$ and $B \to \tau \nu$ processes. First, the effective Lagrangian for $b \to c \tau \nu$ operators is given by
\begin{eqnarray}
\mathcal{L}_{\rm \normalsize eff} &=& -\frac{G_{\rm \normalsize F}}{\sqrt{2}}V_{cb}\overline{c}\gamma_\mu (1 - \gamma_5)b \overline{\tau}\gamma^\mu (1 - \gamma_5)\nu_\tau \nonumber \\
&& + G_{\rm \normalsize S}\overline{c}b \overline{\tau}(1 - \gamma_5)\nu_\tau + G_{\rm \normalsize P}\overline{c}\gamma_5 b \overline{\tau}(1 - \gamma_5)\nu_\tau + {\rm h.c.},
\end{eqnarray}
where $G_{\rm \normalsize S}$ and $G_{\rm \normalsize P}$ are scalar and pseudo-scalar effective couplings. These couplings are given from Eqs.(\ref{down}), (\ref{up}) and (\ref{lepton}),
\begin{eqnarray}
G_{\normalsize \rm S} &\equiv& \frac{\tan^2\beta M_\tau}{2v^2M_{H^\pm}^2}[\hat{R}_e^{-1}]_{33}(M_b[\hat{R}_d^{-1}]_{22}V_{cb} + M_c V_{cb}\cot^2\beta), \label{G_s1} \\
G_{\normalsize \rm P} &\equiv& \frac{\tan^2\beta M_\tau}{2v^2M_{H^\pm}^2}[\hat{R}_e^{-1}]_{33}(M_b[\hat{R}_d^{-1}]_{22}V_{cb} - M_c V_{cb}\cot^2\beta) \label{G_p1}.
\end{eqnarray}
We omit a prime ( $^\prime$ ) from the fields in mass eigenstates. The higgsino diagram contributions to the $[\hat{R}_d^{-1}]_{22}$ is proportional to square of the charm Yukawa couplings. Since the branching ratio can change only by at most a few \%, we neglect such contributions in the followings.
For the case of large $\tan\beta$, we can also neglect the last terms in $G_{\rm S}$ and $G_{\rm P}$.

In order to calculate the $B \to D \tau \nu$ branching ratio, we need vector and scalar form factors of the $B \to D$ transition. In the heavy quark limit, these form factors can be parameterized by a unique function called the Isgur-Wise function. The form of the Isgur-Wise function was investigated by using the dispersion relation \cite{Boyd:1997kz}. From the semi-leptonic decays $B \to D l \nu$ and $B \to D^* l \nu$ ($l = e, \mu$), the Isgur-Wise function is obtained in a one-parameter form, including the short distance and $1 / M_Q$ $(Q = b,c)$ corrections. The short distance corrections for $B \to D \tau \nu$ was also calculated \cite{Miki:2002nz}. Here we adopt the Isgur-Wise function obtained in these literatures, but we do not include the short distance and the $1 / M_Q$ corrections for simplicity. The short distance effects were shown to change the branching ratio within 6\% in the reference of T.Miki et al \cite{Miki:2002nz}.

Using the definitions,
\begin{eqnarray}
x \equiv \frac{2\PB\cdot\PD}{\PB^2},\ \ \ y \equiv \frac{2\PB\cdot\Pt}{\PB^2},\ \ \ \RD \equiv \frac{M_{\normalsize \rm D}^2}{M_{\normalsize \rm B}^2},\ \ \ \Rt \equiv \frac{M_{\normalsize \rm \tau}^2}{M_{\normalsize \rm B}^2}, 
\end{eqnarray}
the differential decay width is given by
\begin{eqnarray}
\frac{d^2\Gamma[B \rightarrow D \tau \nu]}{dxdy} = \frac{G_{\normalsize \rm F}^2|V_{cb}|^2}{128\pi^3}M_{\normalsize \rm B}^5\rho_{\normalsize \rm D}(x,y),
\end{eqnarray}
where
\begin{eqnarray}
\rho_{\normalsize \rm D}(x,y) &\equiv& [|f_+|^2g_1(x,y) + 2{\rm Re}(f_+f_-^{\prime *})g_2(x,y) + |f_-^\prime|^2g_3(x)], \\
g_1(x,y) &\equiv& (3 - x - 2y - \RD + \Rt)(x + 2y - 1 - \RD - \Rt) \nonumber \\
&& - (1 + x + \RD)(1 + \RD - \Rt - x) , \\
g_2(x,y) &\equiv& \Rt(3 - x - 2y - \RD + \Rt) , \\
g_3(x) &\equiv& \Rt(1 + \RD - \Rt - x) , \\
f_-^\prime &\equiv& [f_- -\Delta_{\normalsize \rm S}[f_+(1 - \RD) + f_-(1 + \RD - x)]], \\
f_{\pm} &=& \pm \frac{1 \pm \sqrt{r_{\normalsize \rm D}}}{2 \sqrt[4]{r_{\normalsize \rm D}}}\xi(w),\ \  (w = \frac{x}{2\sqrt{\RD}}). 
\end{eqnarray}
Here $\Delta_{\rm \normalsize S} \equiv \frac{\sqrt{2} G_{\rm \normalsize S} M_{\rm \normalsize B}^2}{G_{\rm \normalsize F} V_{cb} M_{\tau} (M_b - M_c)}$.
We use the following form of the Isgur-Wise function.
\begin{eqnarray}
\xi(w) &=& 1 - 8 \rho^2_1 z + (51 \rho^2_1 - 10) z^2 - (252 \rho^2_1 - 84) z^3 , \\
z &=& \frac{\sqrt{w + 1} - \sqrt{2}}{\sqrt{w + 1} + \sqrt{2}}.
\end{eqnarray}
For the slope parameter, we use $\rho_1^2 = 1.33 \pm 0.22$ \cite{Boyd:1997kz, Miki:2002nz}.

For the $B \to \tau \nu$ process, the relevant four fermion interactions are those of the $b \to u \tau \nu$ type,

\begin{eqnarray}
\mathcal{L}_{\rm \normalsize eff}^\prime &=& -\frac{G_{\normalsize \rm F}}{\sqrt{2}}V_{ub}\overline{u}\gamma_\mu(1 - \gamma_5)b\overline{\tau}\gamma^\mu(1 - \gamma_5)\nu_\tau \nonumber \\
&& + G_{\normalsize \rm S}^\prime\overline{u}b\overline{\tau}(1 - \gamma_5)\nu_\tau + G_{\normalsize \rm P}^\prime\overline{u}\gamma_5b\overline{\tau}(1 - \gamma_5)\nu_\tau + {\rm h.c.}, \\
G_{\normalsize \rm S}^\prime &\equiv& \frac{\tan^2\beta M_\tau}{2v^2M_{H^\pm}^2}[\hat{R}_e^{-1}]_{33}(M_b[\hat{R}_d^{-1}]_{11}V_{ub} + M_u V_{ub}\cot^2\beta), \label{G_s2} \\
G_{\normalsize \rm P}^\prime &\equiv& \frac{\tan^2\beta M_\tau}{2v^2M_{H^\pm}^2}[\hat{R}_e^{-1}]_{33}(M_b[\hat{R}_d^{-1}]_{11}V_{ub} - M_u V_{ub}\cot^2\beta). \label{G_p2}
\end{eqnarray}
Using the matrix elements
\begin{eqnarray}
\langle 0 |\overline{u}\gamma^\mu\gamma_5 b |B^-\rangle &=& i f_{\normalsize \rm B} p^\mu , \\
\langle 0 |\overline{u}\gamma_5 b |B^-\rangle &=& -i f_{\normalsize \rm B} \frac{M_{\normalsize \rm B}^2}{M_b},
\end{eqnarray}
the decay width is given by
\begin{eqnarray}
\Gamma[B \rightarrow \tau \nu] = \frac{G_{\normalsize \rm F}^2}{8\pi}|V_{ub}|^2f_{\normalsize \rm B}^2M_{\normalsize \rm B}^2M_{\normalsize \rm \tau}^2 \left[1 - \frac{2v^2}{M_b M_\tau V_{ub}}G_{\normalsize \rm P}^\prime\right]^2(1 - \Rt)^2,
\end{eqnarray}
where $f_{\rm B}$ is the $B_u$ decay constant.

In the generalized case of the MFV with Eqs(\ref{assumption1}) - (\ref{assumption3}), the scalar and pseudo-scalar couplings, Eqs.(\ref{G_s1}), (\ref{G_p1}), (\ref{G_s2}), and (\ref{G_p2}) can be obtained by the following replacement.
\begin{eqnarray}
{[ \hat{R}_d^{-1} ]}_{22} &\rightarrow& \frac{1}{ 1 + [\Eg^{(3)} - \Egp^{(32)}]\tan\beta}, \label{R_d_22}  \\
{[ \hat{R}_d^{-1} ]}_{11} &\rightarrow& \frac{1}{ 1 + [\Eg^{(3)} - \Egp^{(31)}]\tan\beta}.
\end{eqnarray}
Notice that the right-handed sides of the above equations are approximately same because $\Egp^{(31)} \approx \Egp^{(32)}$. This is the generalization of $[\hat{R}_d^{-1}]_{11} \approx [\hat{R}_d^{-1}]_{22}$, which follows from fact that the higgsino diagram contribution can be neglected in the evaluation with the $[\hat{R}_d^{-1}]_{11}$ and $[\hat{R}_d^{-1}]_{22}$.

\section{Numerical results}
In this section, we shall present results of the numerical calculations on branching ratios of the $B \to D \tau \nu$ and $B \to \tau \nu$ processes in the MSSM. We see that charged Higgs effects to these processes become important for the parameter region of a large $\tan\beta$ and a small charged Higgs mass. We also discuss $B_s \to \mu^+\mu^-$ and $b \to s \gamma$, because SUSY corrections to these processes are important for this parameter region. The relevant SUSY parameters are $\tan\beta$, $M_{H^\pm}$, the higgsino mass parameter $\mu$, the bino mass parameter $M_{\widetilde{B}}$, the gluino mass $M_{\widetilde{g}}$, the sbottom mass $M_{\widetilde{b}}$, and the stau mass $M_{\widetilde{\tau}}$. For the sbottom and the stau, we take the left and right handed masses to be the same, and neglect the left-right mixing terms.

\begin{figure}[h]
\hspace{3cm}\includegraphics[width=.7\linewidth]{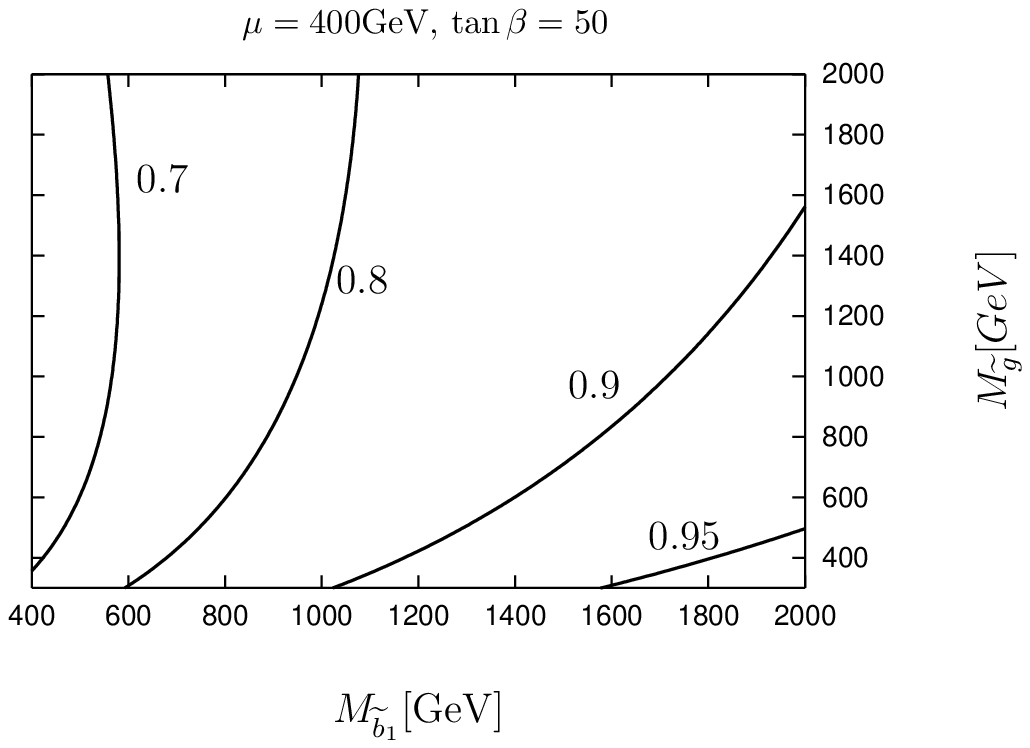}

\hspace{3cm}\includegraphics[width=.7\linewidth]{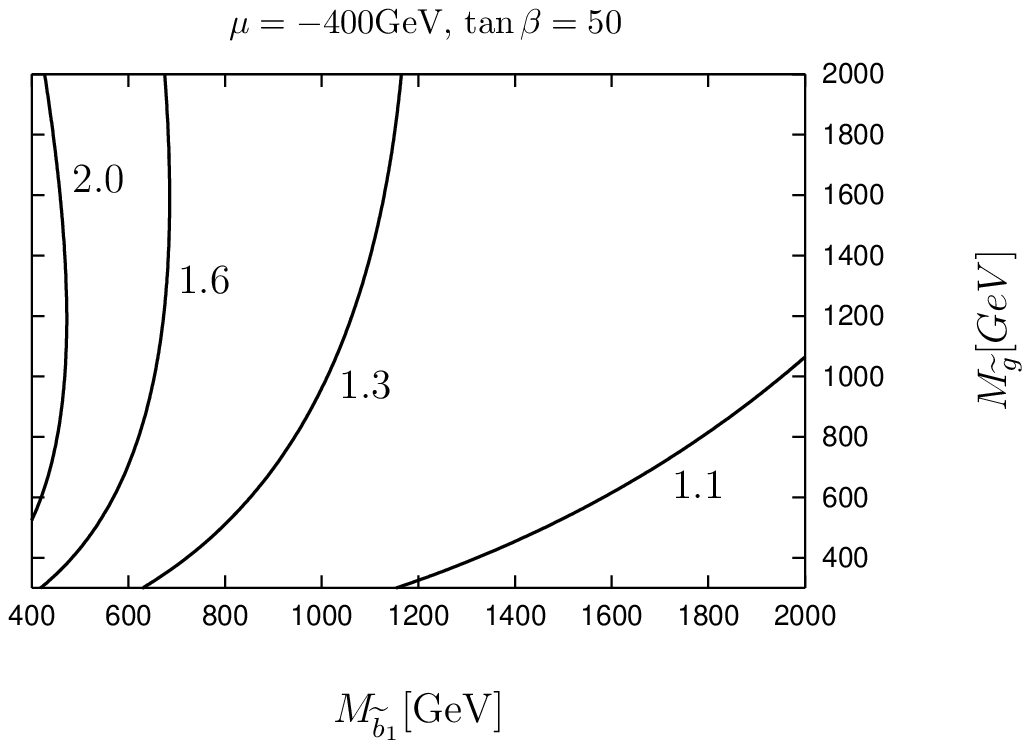}	
\caption{Contour plots of the correction factor $[\hat{R}_d^{-1}]_{ii}(i = 1,2)$ in the $M_{\widetilde{g}}$ and $M_{\widetilde{b}_1}$ space for $\tan\beta = 50$ and $\mu = \pm 400$GeV. The numbers in figure are the values of $[\hat{R}_d^{-1}]_{ii}(i = 1,2)$. The values of $[\hat{R}_d^{-1}]_{ii}(i = 1,2)$ is unity without SUSY corrections.}\label{contourRd}
\end{figure}

\begin{figure}[h]
\hspace{3cm}\includegraphics[width=.7\linewidth]{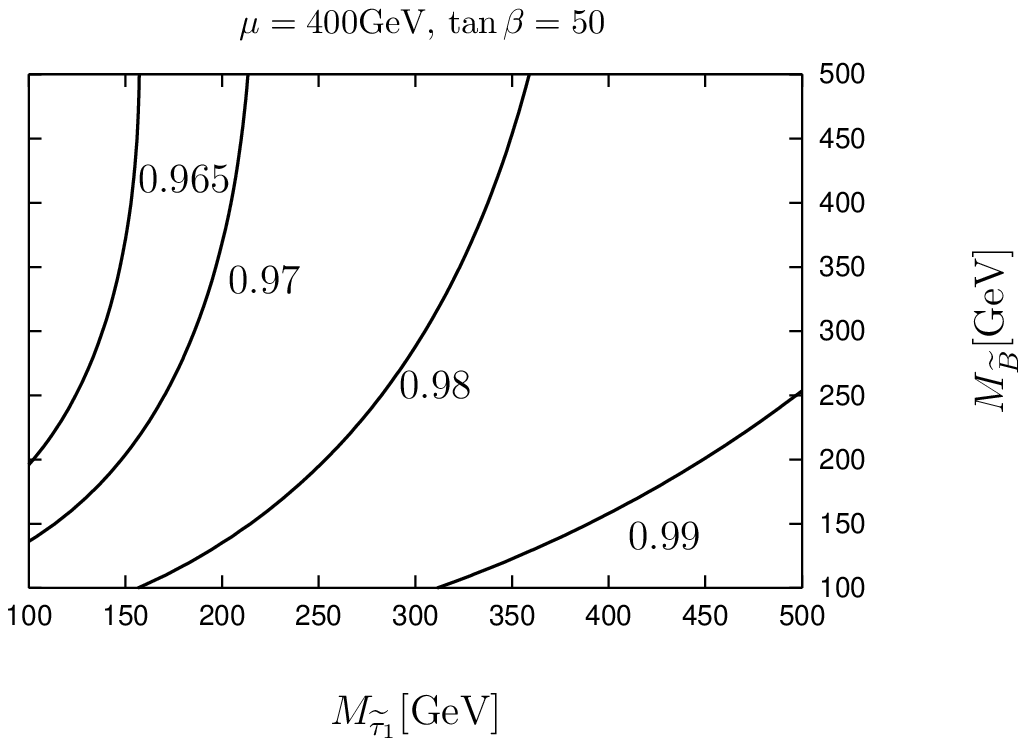}

\hspace{3cm}\includegraphics[width=.7\linewidth]{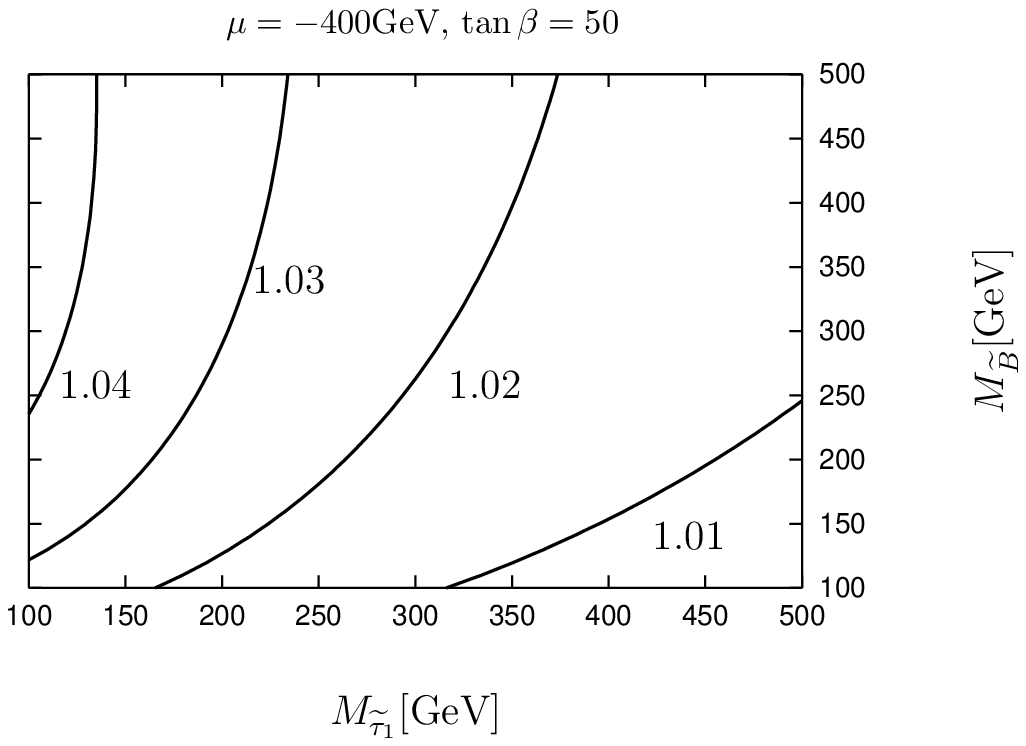}	
\caption{Contour plots of the correction factor $[\hat{R}_e^{-1}]_{33}$ in the $M_{\widetilde{B}}$ and $M_{\widetilde{\tau}_1}$ for $\tan\beta = 50$ and $\mu = \pm 400$GeV. The numbers in figure are the values of $[\hat{R}_e^{-1}]_{33}$. The value of $[\hat{R}_e^{-1}]_{33}$ is unity without SUSY corrections.}\label{contourRe}
\end{figure}

\begin{figure}[h]
\hspace{3cm}\includegraphics[width=.7\linewidth]{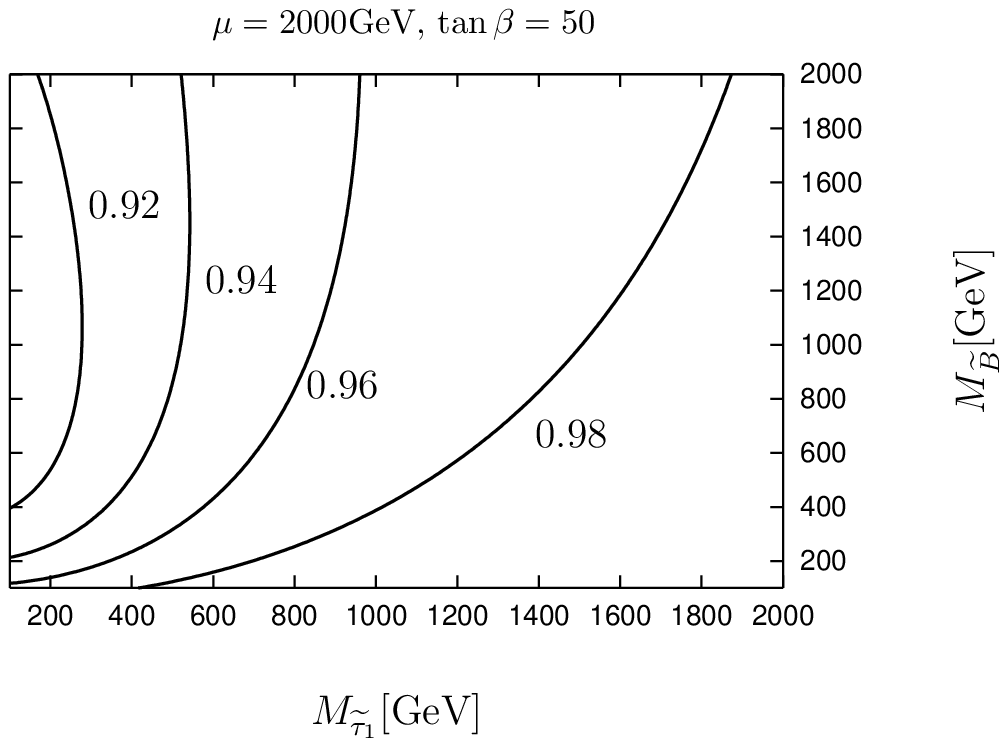}

\hspace{3cm}\includegraphics[width=.7\linewidth]{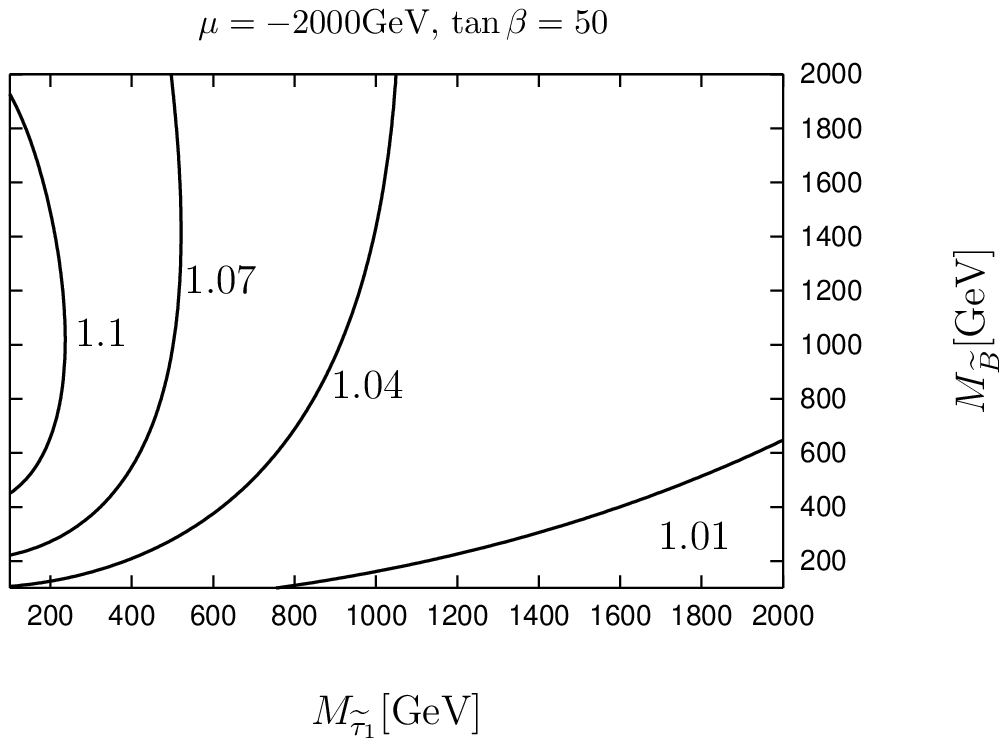}
\caption{Contour plots as Fig.\ref{contourRe} for $\tan\beta = 50$ and $\mu = \pm 2000$GeV. We also change the range of the bino and light stau mass eigenvalue.}\label{contourRe_large}
\end{figure}

We first show $[\hat{R}_d^{-1}]_{22}$ in Eq.(\ref{down}) and $[\hat{R}_e^{-1}]_{33}$ in Eq.(\ref{lepton}). As we discussed in the previous section, $[\hat{R}_d^{-1}]_{11}$ and $[\hat{R}_d^{-1}]_{22}$ are approximately same, because the higgsino loop contributions, Eq.(\ref{A-term dependence}), are suppressed. Therefore the A-term dependence of $[\hat{R}_d^{-1}]_{11,22}$ is negligible. In Fig \ref{contourRd}, contour plots of $[\hat{R}_d^{-1}]_{22}$ are presented in the gluino mass and light sbottom mass eigenvalue $M_{\widetilde{b}_1}$ space in the case of the $\mu = \pm400$GeV and $\tan\beta = 50$. For a positive value of $\mu$, the correction become $-10$ to $-40$ \% in this parameter region. On the other hand, corrections become positive and huge for a negative $\mu$. The value of the $[\hat{R}_e^{-1}]_{33}$ is shown in the light stau mass eigenvalue $M_{\widetilde{\tau}_1}$ and bino mass space for $\mu = \pm400$GeV and $\tan\beta = 50$ in Fig.\ref{contourRe}. In general, the correction is smaller compared with the case of $[\hat{R}_d^{-1}]_{22}$. For a larger value of $\mu$, however the correction can be 10 \%. One such example is shown in Fig.\ref{contourRe_large}.

\begin{figure}[h]
\hspace{2cm}\includegraphics[width=.7\linewidth]{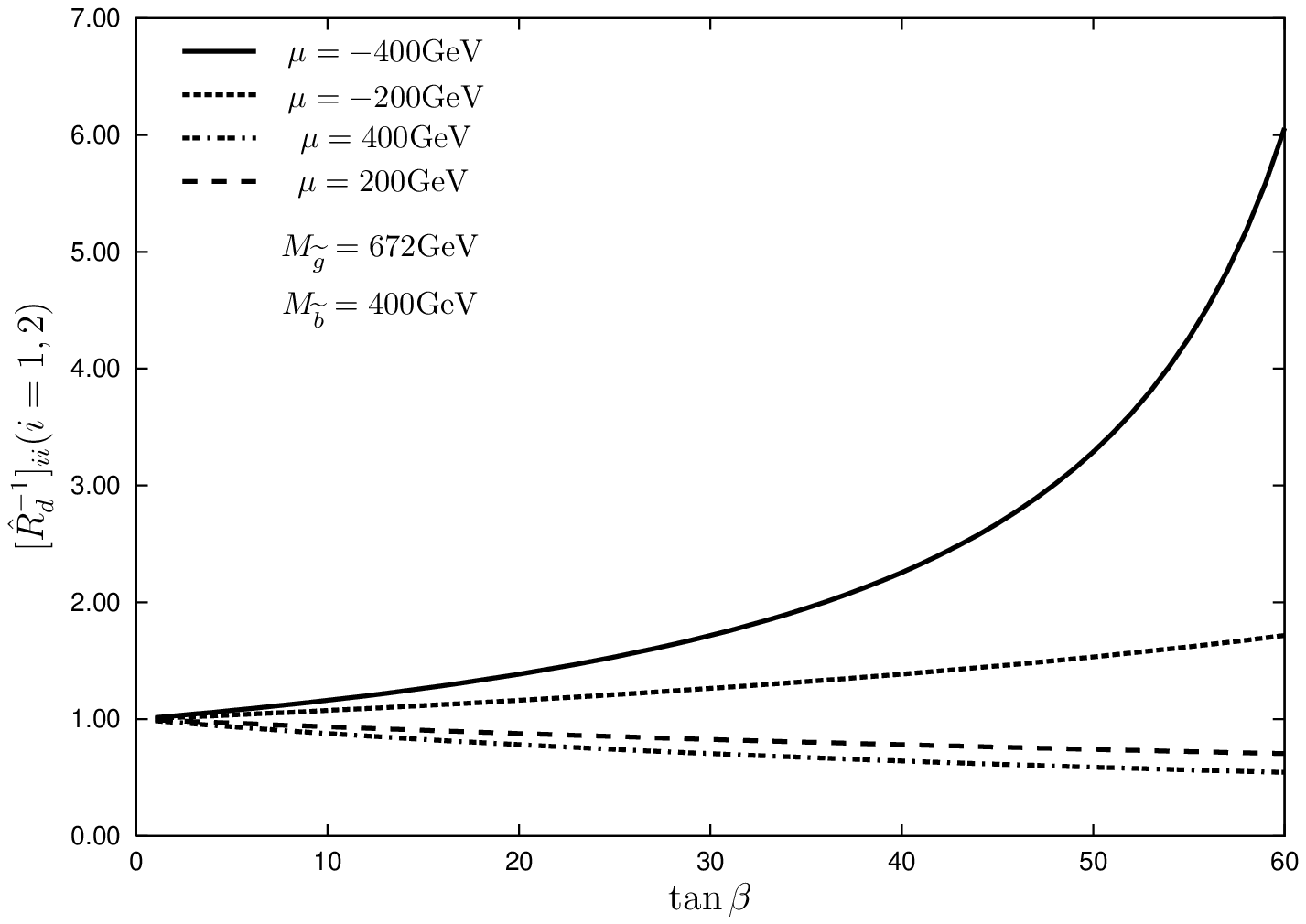}

\hspace{2cm}\includegraphics[width=.7\linewidth]{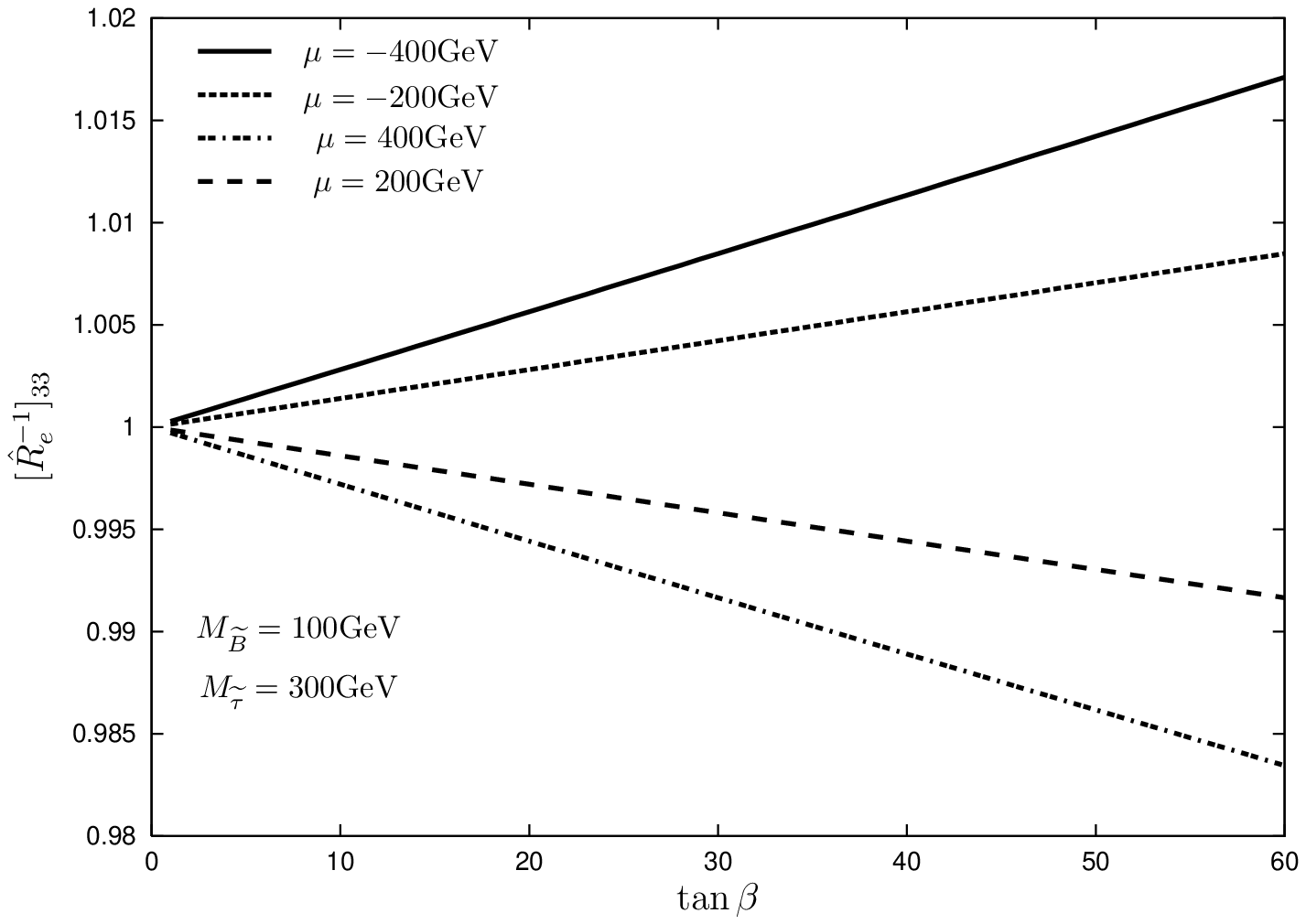}	
\caption{The correction factors as a function of $\tan\beta$. The upper figure is $[\hat{R}_d^{-1}]_{ii}(i = 1,2)$ for $M_{\widetilde{g}} = 672$GeV and $M_{\widetilde{b}} = 400$GeV, and the lower figure is $[\hat{R}_e^{-1}]_{33}$ for $M_{\widetilde{B}} = 100$GeV and $M_{\widetilde{\tau}} = 300$GeV. The value of the $\mu$ parameter is shown in the figures.}\label{correction_tanB}
\end{figure}

The dependence on $\tan\beta$ in $[\hat{R}_d^{-1}]_{22}$ and $[\hat{R}_e^{-1}]_{33}$ are shown Fig.\ref{correction_tanB}. Here we take $\mu = \pm400$GeV and $\mu = \pm200$GeV. The gluino and bino masses satisfy the GUT relation ($M_{\widetilde{g}} = 6.72M_{\widetilde{B}}$). The correction to $[\hat{R}_d^{-1}]_{22}$ becomes large for $\tan\beta \gtrsim 30$.

\begin{figure}[h]
\hspace{2cm}\includegraphics[width=.7\linewidth]{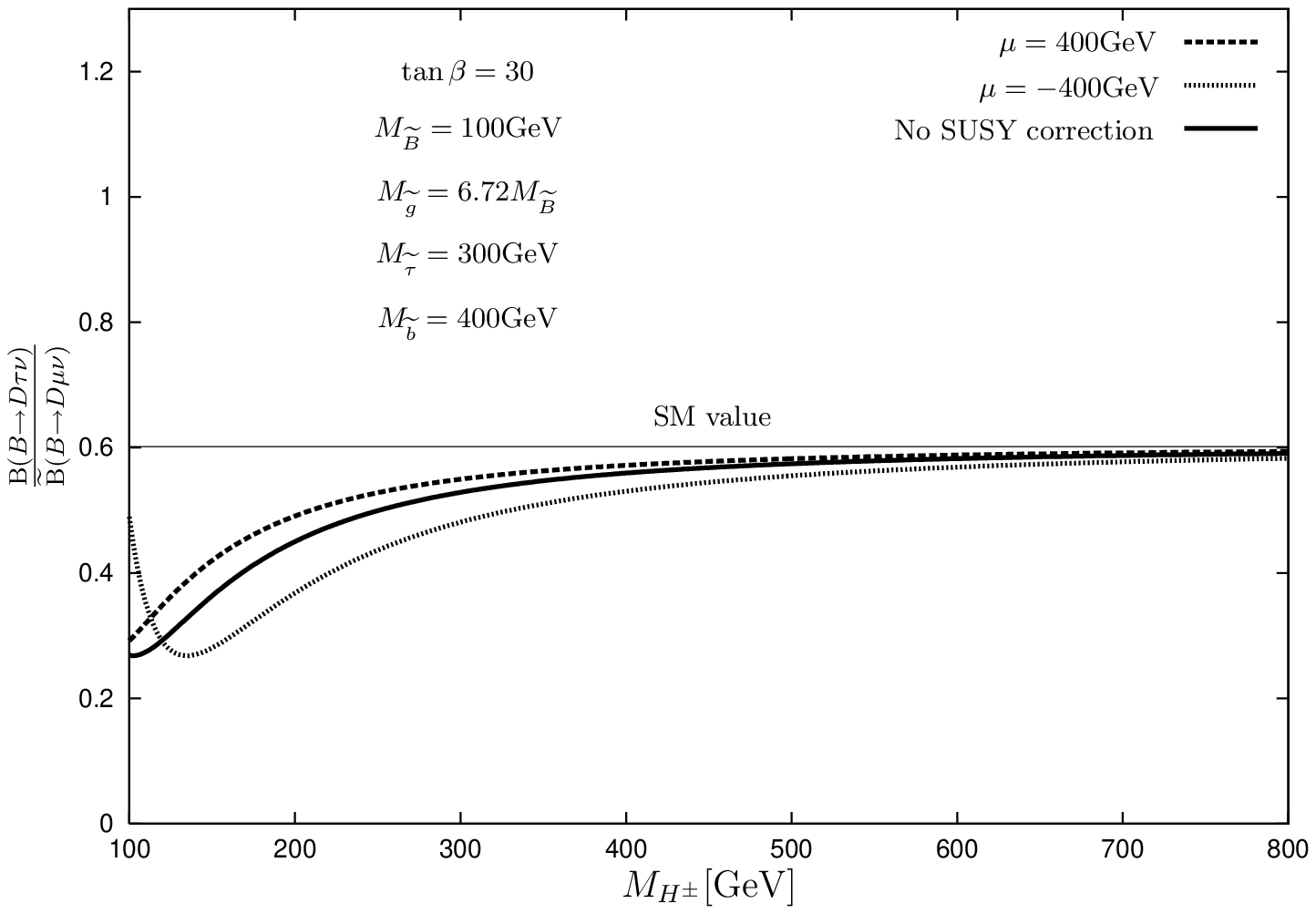}

\hspace{2cm}\includegraphics[width=.7\linewidth]{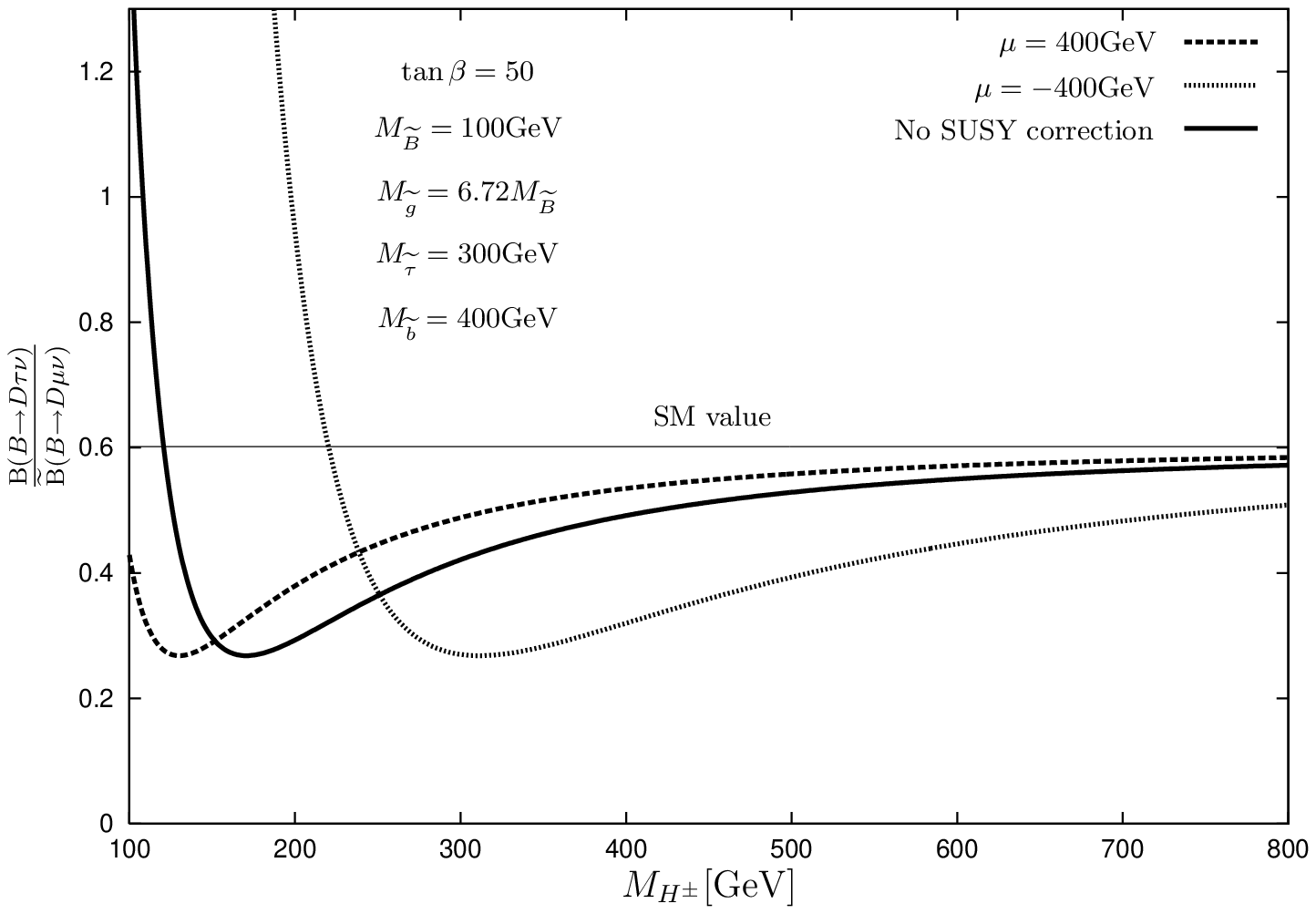}	
\caption{Value of $\frac{{\rm B}(B \to D \tau \nu)}{\widetilde{\rm B}(B \to D \mu \nu)}$ as a function of the charged Higgs mass for $\tan\beta = 30, 50$, $M_{\widetilde{B}} = 100$GeV, $M_{\widetilde{g}} = 6.72M_{\widetilde{B}}$, $M_{\widetilde{b}} = 400$GeV and $M_{\widetilde{\tau}} = 300$GeV. The horizontal solid line is the predicted value in the SM.}\label{BDtn_br}
\end{figure}

\begin{figure}[h]
\hspace{2cm}\includegraphics[width=.7\linewidth]{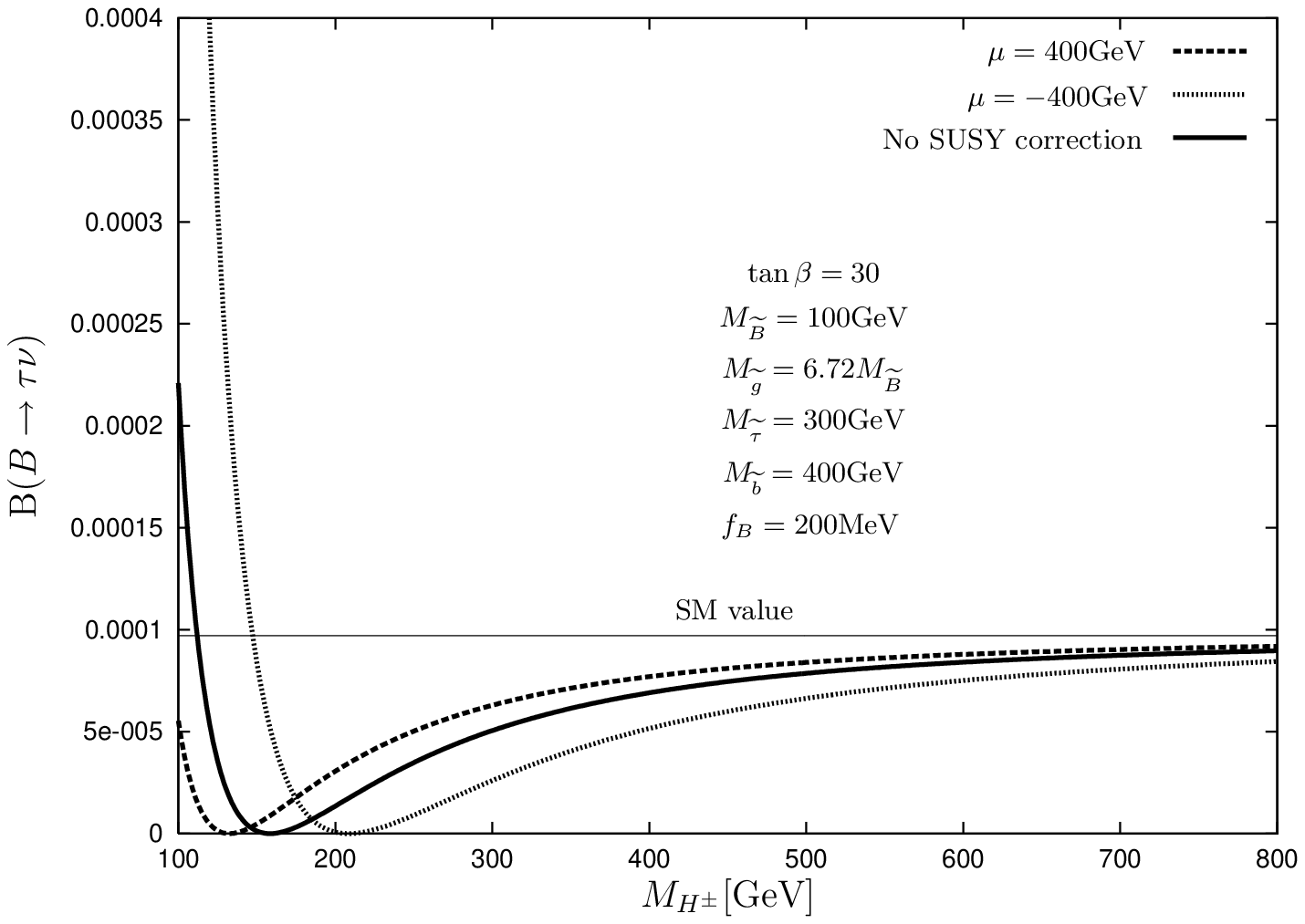}

\hspace{2cm}\includegraphics[width=.7\linewidth]{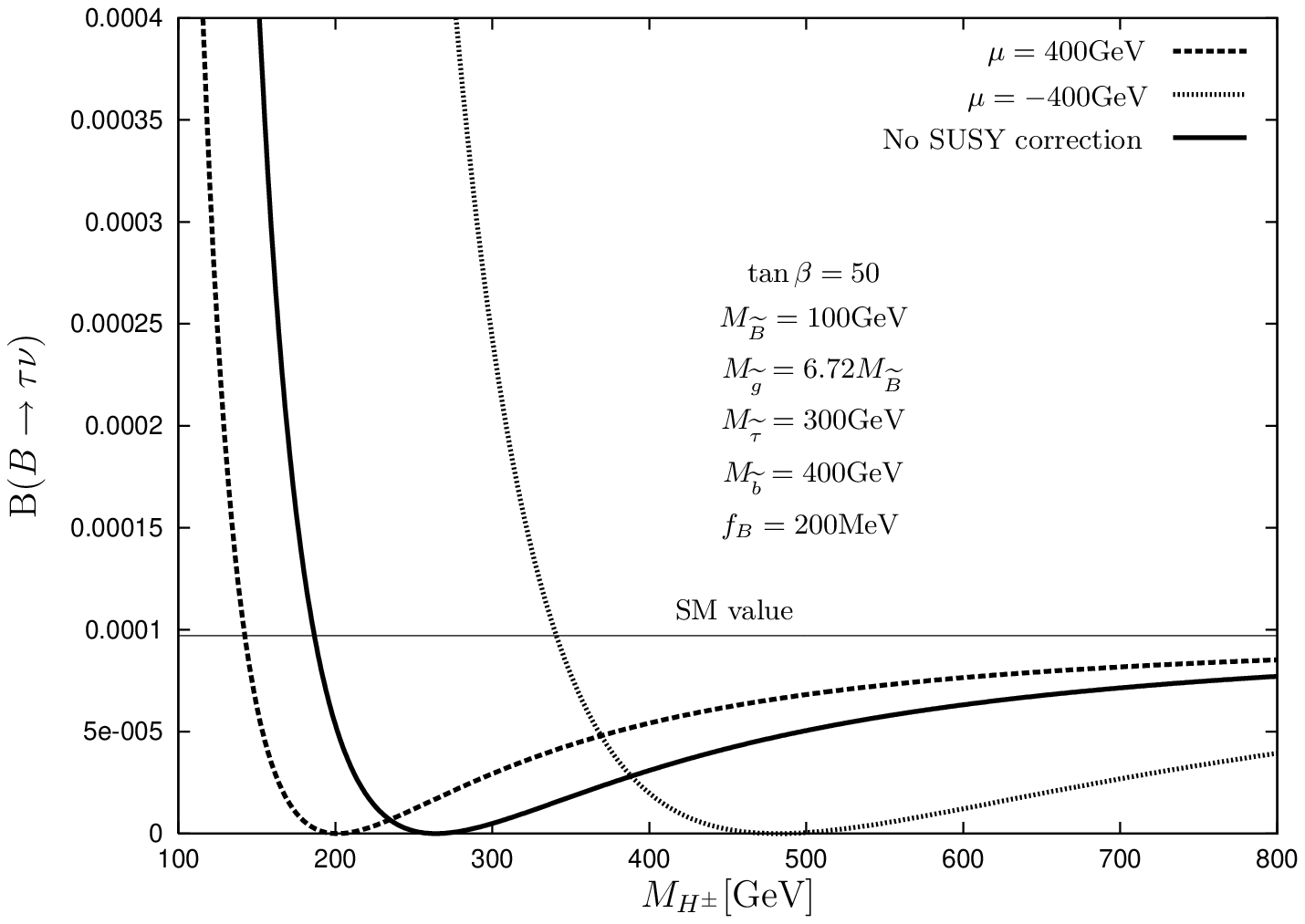}	
\caption{Value of the ${\rm B}(B \to \tau\nu)$ as a function of the charged Higgs mass for $\tan\beta = 30, 50$, $M_{\widetilde{B}} = 100$GeV, $M_{\widetilde{g}} = 6.72M_{\widetilde{B}}$, $M_{\widetilde{b}} = 400$GeV, $M_{\widetilde{\tau}} = 300$GeV and $f_B = 200$MeV. The horizontal solid line is the predicted value in the SM.}\label{Btn_br}
\end{figure}

Next, we show the branching ratio of $B \to D \tau \nu$ and $B \to \tau \nu$. For $B \to D \tau \nu$, we consider the following ratio of the two branching ratios, where $\widetilde{\rm B}(B \to D \mu \nu)$ is defined as the branching fraction for the $B \to D \mu \nu$ mode integrated over the same phase space of the $B \to D \tau \nu$ kinematics.

\begin{eqnarray}
\frac{{\rm B}(B \to D \tau \nu)}{{\rm \widetilde{B}}(B \to D \mu \nu)} = \frac{\int^{1 + r_D - r_\tau}_{2\sqrt{r_D}}dx \int^{y_2}_{y_1}dy \frac{G_{\rm \normalsize F}^2 |V_{cb}|^2}{128\pi^3}M_{\rm \normalsize B}^5 \rho_D(x,y) }{\int^{1 + r_D - r_\tau}_{2\sqrt{r_D}}dx \int^{y_2^\prime}_{y_1^\prime}dy \frac{G_{\rm \normalsize F}^2 |V_{cb}|^2}{128\pi^3}M_{\rm \normalsize B}^5 \rho_D^\prime(x,y)},
\end{eqnarray}
\begin{eqnarray}
y_{1,2} &\equiv& \frac{\left(1 - \frac{x}{2} \pm \sqrt{\frac{x^2}{4} - r_D}\right)^2 + r_\tau}{1 - \frac{x}{2} \pm \sqrt{\frac{x^2}{4} - r_D}}, \\
y_{1,2}^\prime &\equiv& \frac{\left(1 - \frac{x}{2} \pm \sqrt{\frac{x^2}{4} - r_D}\right)^2 + r_\mu}{1 - \frac{x}{2} \pm \sqrt{\frac{x^2}{4} - r_D}}.
\end{eqnarray}
For the $B \to D \mu \nu$ mode, we use
\begin{eqnarray}
\rho_{\normalsize \rm D}^\prime(x,y) &\equiv& [|f_+|^2g_1^\prime(x,y) + 2{\rm Re}(f_+f_-^{\prime *})g_2^\prime(x,y) + |f_-^\prime|^2g_3^\prime(x)], \\
g_1^\prime(x,y) &\equiv& (3 - x - 2y - \RD + r_\mu)(x + 2y - 1 - \RD - r_\mu) \nonumber \\
&& - (1 + x + \RD)(1 + \RD - r_\mu - x) , \\
g_2^\prime(x,y) &\equiv& r_\mu(3 - x - 2y - \RD + r_\mu) , \\
g_3^\prime(x) &\equiv& r_\mu(1 + \RD - r_\mu - x) , \\
r_\mu &\equiv& \frac{M_\mu^2}{M_{\rm \normalsize B}^2}.
\end{eqnarray}
It was pointed out that we can reduce theoretical uncertainty associated with form factors by taking this ratio \cite{Tanaka:1994ay}. In Fig.\ref{BDtn_br}, we show the above quantity as a function of the charged Higgs mass for the parameter set of $M_{\widetilde{b}} = 400$GeV, $M_{\widetilde{B}} = 100$GeV, $M_{\widetilde{\tau}} = 300$GeV and $\tan\beta = 30$ and $50$. In this figure, we also draw a line without the SUSY corrections. We can see a large deviation due to the SUSY loop corrections. A similar figure for $B \to \tau \nu$ are shown in Fig.\ref{Btn_br}. We can see that ${\rm B}(B \to \tau \nu)$ vanishes in a particular point of $M_{H^\pm}$ depending on SUSY parameters, and below that point the branching ratio is significantly enhanced.

We consider the correlation between the branching ratio of the $B \to D \tau \nu$ and that of the $B \to \tau \nu$. Under the assumption of MFV, the charged Higgs effect appears through the following combination of the parameters in the branching ratio formulas,
\begin{eqnarray}
\widetilde{R} \equiv \frac{M_W \tan\beta}{M_{H^\pm}}\sqrt{[\hat{R}_e^{-1}]_{33}[\hat{R}_d^{-1}]_{22}}.\label{R_tilde}
\end{eqnarray}
(or the replacement of $[\hat{R}_d^{-1}]_{22}$ by the right hand side of Eq.(\ref{R_d_22}) in the generalized MFV case.)
This is compared with the type I{}I 2HDM, where $\widetilde{R}$ is replaced by $R$,
\begin{eqnarray}
R \equiv \frac{M_W \tan\beta}{M_{H^\pm}}.
\end{eqnarray} 
In other words, the SUSY corrections effectively change the value of the $\tan\beta$ in the formula of the 2HDM. Therefore the correlation between two branching ratios is the same for MSSM and 2HDM. Note that this correlation arises due to our assumption of the MFV, namely from the fact that the higgsino contribution does not induce sizable effects in these branching ratios. 
\begin{figure}[h]
\hspace{2cm}\includegraphics[width=.7\linewidth]{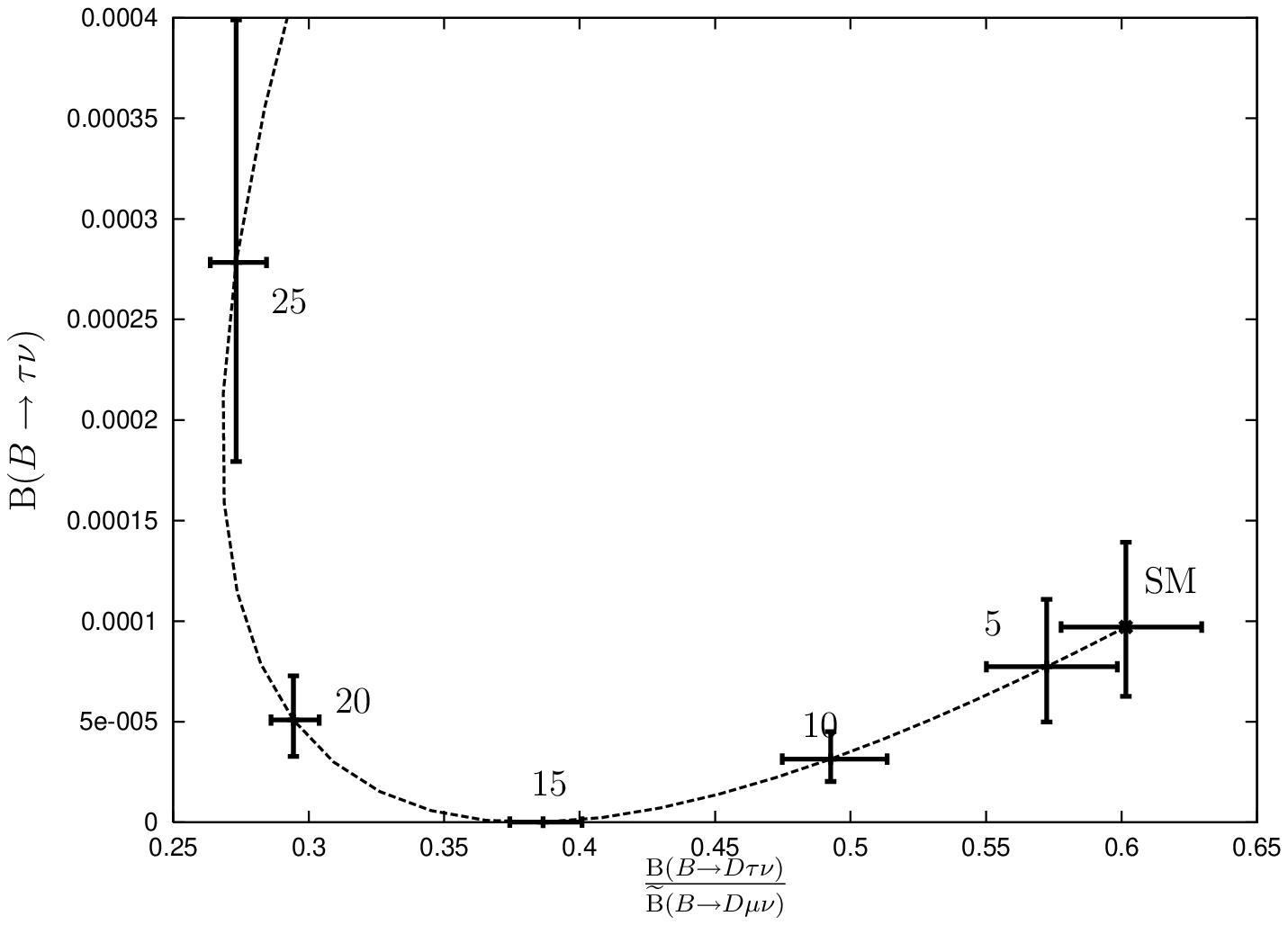}
\caption{Correlation between $\frac{{\rm B}(B \to D \tau \nu)}{\widetilde{\rm B}(B \to D \mu \nu)}$ and ${\rm B}(B \to \tau\nu)$ for the MSSM with MFV. The values on the line represent $\widetilde{R}$ defined in Eq.(\ref{R_tilde}). The theoretical uncertainties associated with $\rho_1^2$ for $B \to D \tau \nu$/$B \to D \mu \nu$ and $f_B$ and $|V_{ub}|$ for $B \to \tau \nu$ are also shown for each point.}\label{correlation}
\end{figure}
In Fig.\ref{correlation}, we show the correlation of the two quantities. We also show estimated theoretical uncertainties for several values of $\widetilde{R}$ along the line. The error corresponds to uncertainty from $\rho_1^2 = 1.33 \pm 0.22$ for $B \to D \tau \nu$, and from $f_B = 200 \pm 30$MeV and $|V_{ub}| = (3.67 \pm 0.47) \times 10^{-3,}$ \cite{Eidelman:2004wy} for $B \to \tau \nu$. We see that uncertainty of $B(B \to \tau \nu)$ from the present input parameters is still large. These errors, however, can be reduced significantly in future from more precise determination of semi-leptonic B decay form factors and $|V_{ub}|$, and improvement of $f_B$ determination from the lattice gauge theory. For instance, if the uncertainty of the slope parameter $\rho_1^2$ is improved to $\pm 0.10$, the theoretical uncertainty from this source to $\rm{B}(B \to D \tau \nu)$ is reduced from $\pm 5 \%$ to $\pm 2 \%$ for the SM case. If the uncertainty of $|V_{ub}|$ is improved to $\pm 5.8\%$, the error of $\rm{B}(B \to \tau \nu)$ is reduced from $\pm 39 \%$ to $\pm 32 \%$, and further improvement of $|V_{ub}|$ and $f_B$ to $\pm 4.4 \%$ and $\pm 16\rm{MeV}$ respectively leads to $\pm 18 \%$ \cite{L.O.I.}. At the SuperKEKB, it is expected that the sensitivity to $\widetilde{R}$ reaches to $11$ (90\% confidence level) for an integrated luminosity of $5 {\rm ab}^{-1}$ from the $B \to D \tau \nu$ process \cite{L.O.I.}. Observation of the $B \to \tau \nu$ mode is possible at 30${\rm ab}^{-1}$ for the SM case \cite{S.Nishida}.

Let us comment on scaling behaviors of the SUSY loop corrections. $\hat{\bf E}_{\widetilde{g}}$, $\hat{\bf E}_{\widetilde{h}}$, and $\hat{\bf E}_{\widetilde{B}}$ in Eqs.(\ref{gluino loop}), (\ref{higgsino loop}), and (\ref{bino loop}) remain constants when all SUSY mass parameters are multiplied by a same factor. Therefore the SUSY loop effects to the charged Higgs contribution in $B \to D \tau \nu$ and $B \to \tau \nu$ do not decouple by taking large SUSY mass spectrum as long as the charged Higgs mass is the same. This situation is similar to the SUSY loop contributions to the neutral Higgs exchange in the flavor changing neutral current and lepton flavor violation processes.\cite{FCNC, Dedes:2002er, Huang:1998vb, Babu:2002et, Dedes:2002rh, Sher:2002ew, Kitano:2003wn}

\begin{figure}[h]
\hspace{2cm}\includegraphics[width=.7\linewidth]{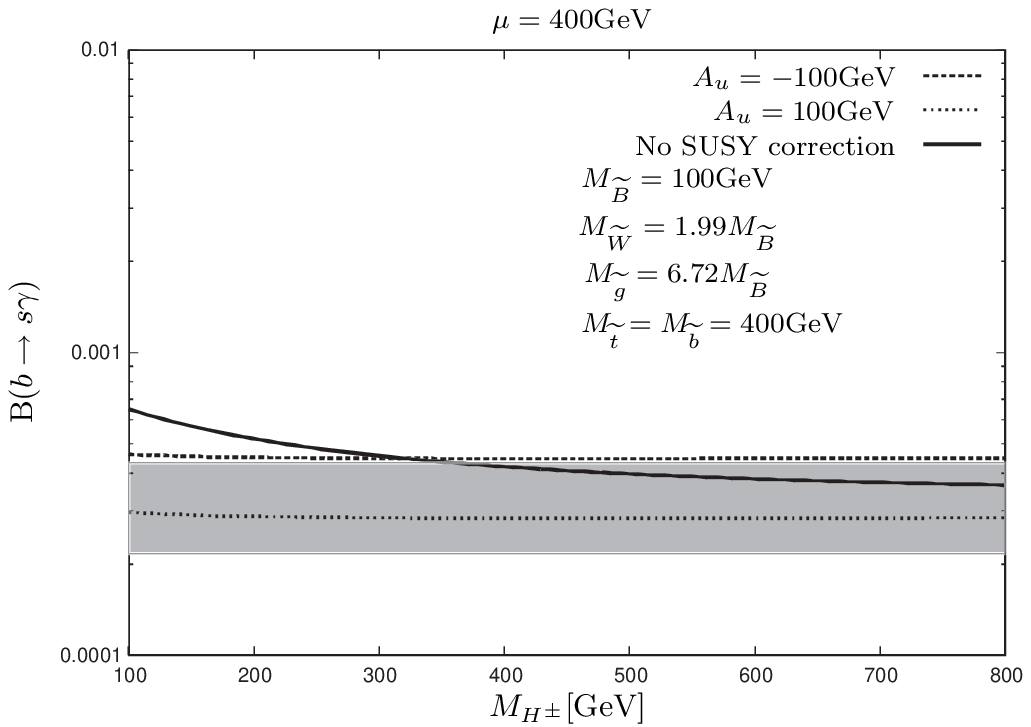}

\hspace{2cm}\includegraphics[width=.7\linewidth]{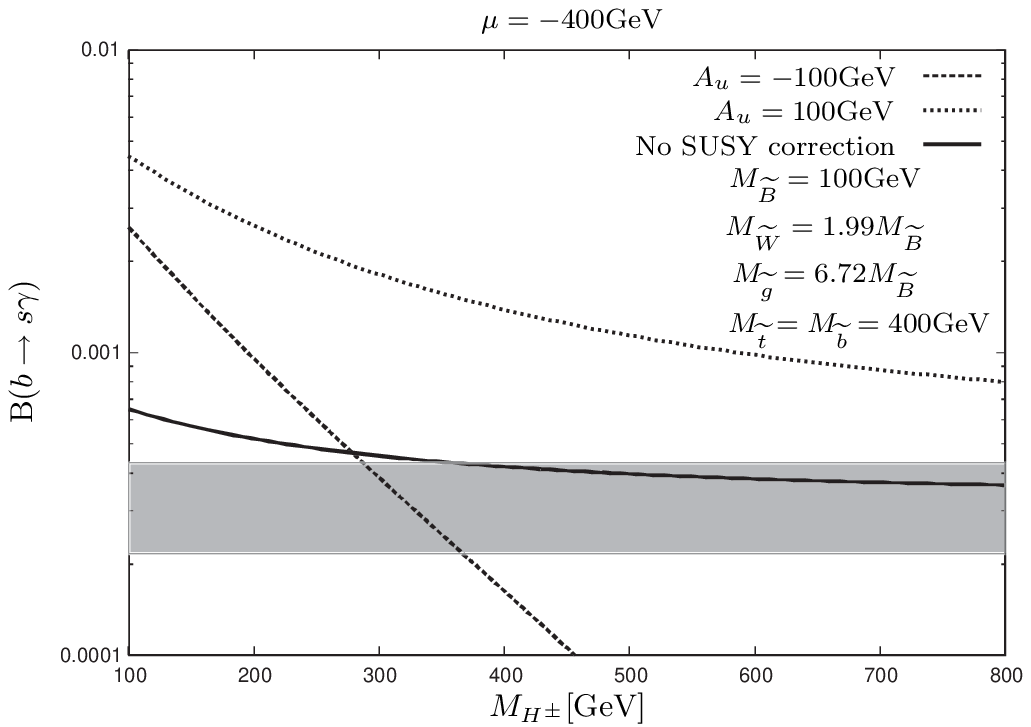}	
\caption{$b \to s \gamma$ branching ratio as a function of the charged Higgs mass for $M_{\widetilde{B}} = 100$GeV, $M_{\widetilde{W}} = 1.99M_{\widetilde{B}}$, $M_{\widetilde{g}} = 6.72M_{\widetilde{B}}$, $M_{\widetilde{t}} = M_{\widetilde{b}} = 400$GeV, $|A_u| = 100$GeV, $|\mu| = 400$GeV, and $\tan\beta = 50$. The shaded region is the experimental allowed region at $2\sigma$ level. \cite{Abe:2001hk}. }\label{b_sgamma}
\end{figure}

Finally, we discuss correlation of the tauonic B decays with $b \to s \gamma$ and $B_s \to \mu^+\mu^-$. It is known that these processes receive significant SUSY contributions for the large $\tan\beta$ case. In the MFV, the $b \to s \gamma$ amplitude consists of the SM contribution, the charged Higgs contribution, and the chargino-stop contribution. The gluino-sbottom contribution is not significant for MFV. The effect of SUSY loop correction to the charged Higgs vertex was also studied \cite{Degrassi:2000qf,Carena:2000uj}. We calculate the $b \to s \gamma$ branching ratios following the formula presented in G.Degrassi et al \cite{Degrassi:2000qf}. Since the $b \to s \gamma$ process depends on the chargino-stop diagram, there is no strict correlation between $b \to s \gamma$ and tauonic B decays. However, the contribution from the charged Higgs diagrams is enhanced for $\mu < 0$ from the correction by $\hat{R}_d^{-1}$ in Eq.(\ref{down}). As an example, we show ${\rm B}(b \to s \gamma)$ for the following parameter sets, $M_{\widetilde{B}} = 100$GeV, the wino mass $M_{\widetilde{W}} = 1.99M_{\widetilde{B}}$, $M_{\widetilde{g}} = 6.72M_{\widetilde{B}}$, $M_{\widetilde{t}} = M_{\widetilde{b}} = 400$GeV, $|A_u| = 100$GeV, $|\mu| = 400$GeV, and $\tan\beta = 50$ in Fig.\ref{b_sgamma}. In order to satisfy the experimental constraint, a fine tuning between the charged Higgs and chargino contributions is necessary for $\mu < 0$. For $\mu > 0$, the constraint is generally weak.

\begin{figure}[h]
\hspace{2cm}\includegraphics[width=.7\linewidth]{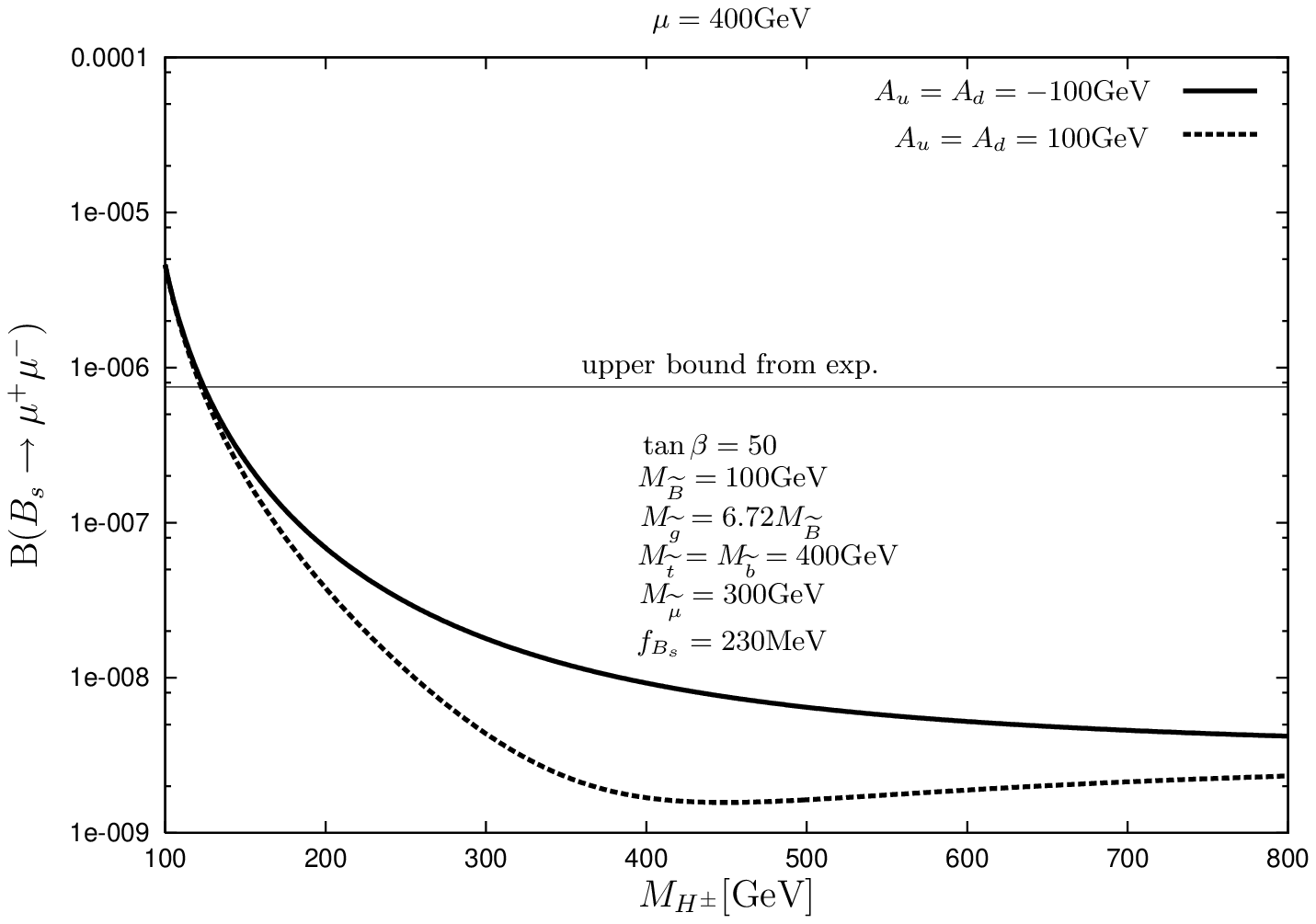}

\hspace{2cm}\includegraphics[width=.7\linewidth]{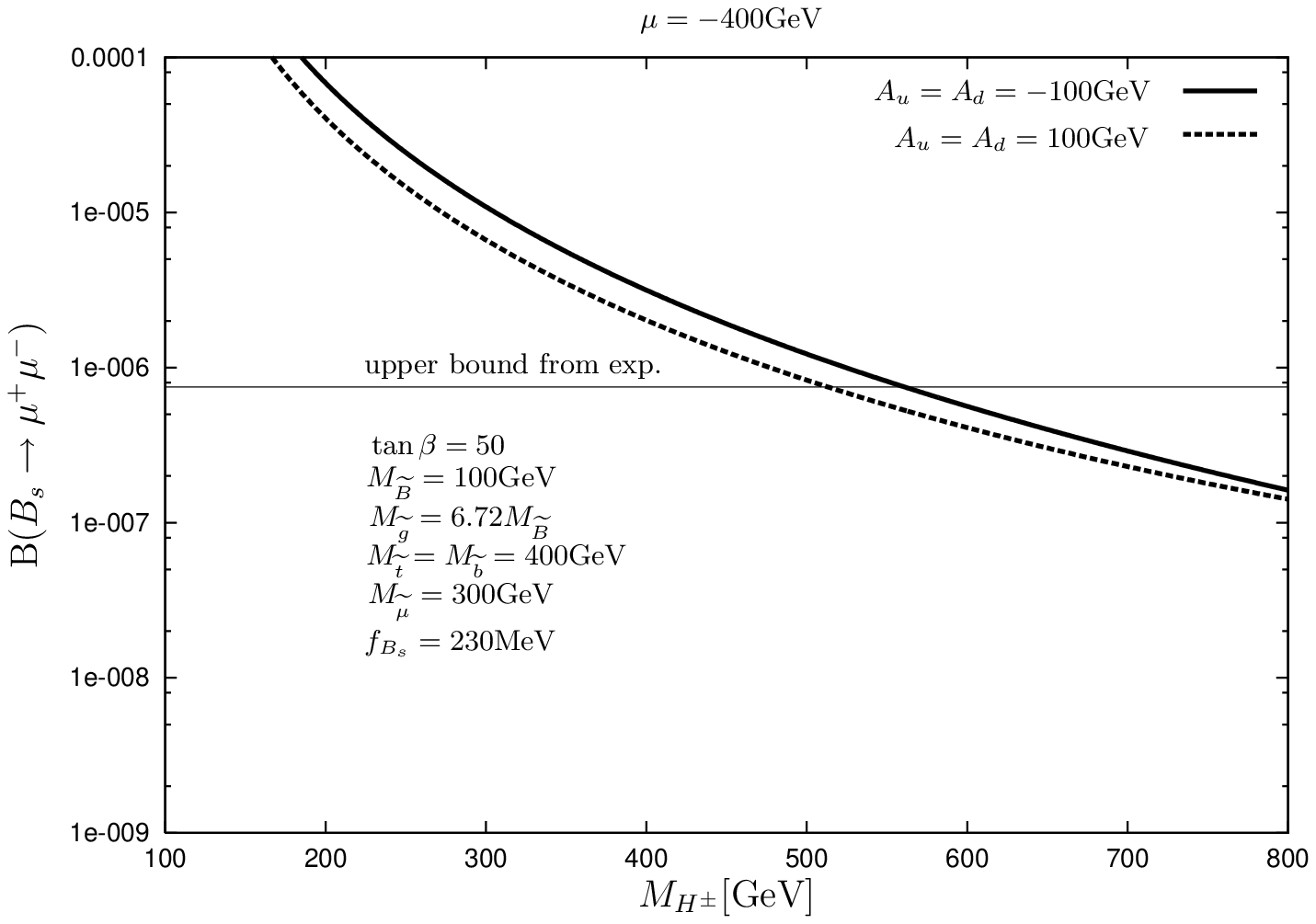}
\caption{Branching ratio of the $B_s \to \mu^+\mu^-$ as a function of the charged Higgs mass for $\tan\beta = 50$, $|A_u| = |A_d| = 100$GeV, $|\mu| = 400$GeV, $M_{\widetilde{t}} = M_{\widetilde{b}} = 400$GeV, $M_{\widetilde{B}} = 100$GeV, $M_{\widetilde{g}} = 6.72M_{\widetilde{B}}$, $M_{\widetilde{\mu}} = 300$GeV, and $f_{B_s} = 230$MeV. The horizontal solid line is the current experimental upper bound \cite{Acosta:2004xj}.}\label{Bs_mumu}
\end{figure}

We also calculate the branching ratio of the $B_s \to \mu^+\mu^-$ in this model. In this case, the branching ratio depends on the higgsino-stop diagram and the gluino-sbottom diagram. In particular, inclusion of the higgsino loop contribution is necessary to generate this flavor changing process. We present the branching ratio of B$(B_s \to \mu^+\mu^-)$ in Fig.\ref{Bs_mumu} for the parameter sets, $\tan\beta = 50$, $|A_u| = |A_d| = 100$GeV, $|\mu| = 400$GeV, $M_{\widetilde{t}} = M_{\widetilde{b}} = 400$GeV, $M_{\widetilde{B}} = 100$GeV, $M_{\widetilde{g}} = 6.72M_{\widetilde{B}}$, $M_{\widetilde{\mu}} = 300$GeV, and $f_{B_s} = 230$MeV. The present experimental upper bound, $7.5 \times 10^{-7,}$, \cite{Acosta:2004xj} is also shown. We can see that the large branching ratio is expected for the parameter space where the tauonic B decays receive significant contribution from the charged Higgs diagram. The enhancement is particularly large for $\mu < 0$. 

The SUSY loop corrections can be also important in the anomalous magnetic moment of muons. In this case the relevant SUSY parameters are the slepton masses, gaugino masses, $\mu$, and $\tan\beta$. For example, the parameter set taken in Fig.\ref{b_sgamma} and \ref{Bs_mumu}, the SUSY contributions to $(g-2)_\mu/2$ is about $7 \times 10^{-9}$, which is larger than the current discrepancy between the experimental result and SM prediction.\cite{Davier:2003pw} If the slepton mass is taken to be larger than 400GeV, the SUSY contributions to $(g-2)_\mu/2$ can be compatible with the current discrepancy. Notice that slepton contribution to the $B_s \to \mu^+ \mu^-$ and tauonic B decays are sub-dominant and that for $b \to s \gamma$ is negligible. Therefore the SUSY contributions to $B_s \to \mu^+ \mu^-$ and tauonic B decays can be important while both $(g-2)_\mu/2$ and $b \to s\gamma$ constraints are satisfied.

\section{Conclusions}
In this paper we study SUSY effects on the tauonic B decays, $B \to D \tau \nu$ and $B \to \tau \nu$ under the assumption of the MFV. These processes receive large corrections for large $\tan\beta$ regime through SUSY loop diagrams related to the bottom and tau Yukawa couplings. For the bottom Yukawa coupling, we find that only the gluino-sbottom loop can contribute significantly and the chargino loop contribution is at a few \% level. This is in contrast to the case of the correction to the $b$-$t$-$H$ vertex, where the stop-chargino loop correction is also important \cite{Carena:1999py}. The effect of the stau-neutralino loop to the tau vertex is generally not as significant as the sbottom-gluino loop, but in some parameter space these contributions can change the charged Higgs exchange effect by more than 10 \%. We also study the correlation between B$(B \to D \tau \nu)$ and B$(B \to \tau \nu)$ within the assumption of the MFV. The SUSY effect on these processes can be absorbed as an effective change of the $\tan\beta$ value, so that the correlation itself is the same as the 2HDM without SUSY loops. It would be therefore interesting to compare this value with $\tan\beta$ measured from other processes in MSSM such as the heavy Higgs direct production \cite{Feng:1996xv}, the chargino/neutralino mixing \cite{Feng:1995zd}, and the stau decay \cite{Nojiri:1996fp}. SUSY corrections to the Higgs couplings can also change the light-Higgs branching ratios \cite{Babu:1998er} and the search limit of the SUSY Higgs bosons at LHC. From combined analysis of Super B Factory and collider experiments, we may be able to obtain important insight on flavor mixing for squarks and sleptons. For example in the present case of MFV, SUSY corrections to $b$-$c$-$H$ and $b$-$u$-$H$ verteces are the same while that to the $b$-$t$-$H$ vertex can be different because of the higgsino loop contribution.

We also compare the tauonic B decay fractions with $b \to s \gamma$ and $B_s \to \mu^+\mu^-$ branching ratios. In general, these processes receive large corrections when we expect large effects in tauonic B decays, i.e. a large $\tan\beta$ and small $M_{H^\pm}$ region. Since stop and chargino diagrams are essential in these flavor changing neutral current processes, we do not have a strict correlation among these processes. However the parameter space with $\mu < 0$ is strongly constrained by $b \to s \gamma$ and $B_s \to \mu^+\mu^-$.

The tauonic B decay processes considered here provide important information on the Yukawa interaction associated with the charged Higgs boson. A large deviation from the SM prediction is expected for large $\tan\beta$ cases from both tree level and SUSY loop effects. The tauonic B decays at future B factory experiments therefore can play a unique role in exploring SUSY models.

\section*{Acknowledgements}
We wish to thank M.E.Peskin and M.Tanaka for useful comments. The work of S.K. was supported by the JSPS Research Fellowships for Young Scientists. The work of Y.O. was supported by a Grant-in-Aid of the Ministry of Education, Culture, Sports, Science and Technology, Government of Japan, Nos. 13640309, 13135225, and 16081211.

\appendix
\section{Charged Higgs coupling for the general MFV case} 
In section 2, we have derived the resummed effective Lagrangian for the charged Higgs couplings under the assumption that soft SUSY breaking mass matrices for the squarks are proportional to a unit matrix in the flavor space. In this appendix, we relax this assumption and obtain the charged Higgs couplings in a more general case of the MFV. General consideration on the MFV case is given in the literature. \cite{D'Ambrosio:2002ex, Buras:2002vd}

We consider the following squark mass matrices and A-terms. This form is motivated from the renormalization group effects on the squark mass matrices in the SUSY breaking scenarios such as minimal supergravity, gauge mediation, and anomaly mediation.
\begin{eqnarray}
M_{{\widetilde{Q}_{\rm \normalsize L}}}^2 &=& [a_1 {\bf 1} + b_1 {\bf y}_u^\dag {\bf y}_u + b_2 {\bf y}_d^\dag {\bf y}_d]\widetilde{M}^2, \\
M_{{\widetilde{U}_{\rm \normalsize R}}}^2 &=& [a_2 {\bf 1} + b_3 {\bf y}_u {\bf y}_u^\dag]\widetilde{M}^2, \\
M_{{\widetilde{D}_{\rm \normalsize R}}}^2 &=& [a_3 {\bf 1} + b_4 {\bf y}_d {\bf y}_d^\dag]\widetilde{M}^2, \\
{\bf A}_{u ij} &=& A_u {\bf y}_{u ij}, \\
{\bf A}_{d ij} &=& A_d {\bf y}_{d ij}, 
\end{eqnarray}
where $a_{1,2,3}$ and $b_{1,2,3,4}$ are real parameters. In the following we take the basis where the down-type Yukawa coupling is diagonal.
\begin{eqnarray}
{\bf y}_u &=& \hat{\bf y}_u \KMtree , \\
{\bf y}_d &=& \hat{\bf y}_d,
\end{eqnarray}
where $\KMtree$ is the flavor mixing matrix in the original Yukawa coupling. Furthermore when we calculate the loop diagrams, we use approximation that only the top Yukawa coupling is kept in the up-type Yukawa coupling.
\begin{eqnarray}
\hat{\bf y}_u \approx \left( \begin{array}{ccc}
							0&& \\
							&0& \\
							&&y_t \\
							\end{array} \right).
\end{eqnarray}
Explicit form of the mass matrix is given by
\begin{eqnarray}
M_{\widetilde{Q}_{\rm \normalsize L} ij}^2 &\approx&  \left( \begin{array}{ccc}
				[a_1 + b_2 y_d^2] \widetilde{M}^2 && \\
				&[a_1 + b_2 y_s^2] \widetilde{M}^2 & \\
				&& [a_1 + b_1 y_t^2 + b_2 y_b^2] \widetilde{M}^2
			\end{array} \right) \\ 
			&& + \left( \begin{array}{ccc}
						& b_1 y_t^2 \KMtreedag_{13}\KMtree_{32}\widetilde{M}^2 &  b_1 y_t^2 \KMtreedag_{13}\KMtree_{33}\widetilde{M}^2 \\
						b_1 y_t^2 \KMtreedag_{23}\KMtree_{31}\widetilde{M}^2 && b_1 y_t^2 \KMtreedag_{23}\KMtree_{33}\widetilde{M}^2 \\
						b_1 y_t^2 \KMtreedag_{33}\KMtree_{31}\widetilde{M}^2 & b_1 y_t^2 \KMtreedag_{33}\KMtree_{32}\widetilde{M}^2 & \\
					\end{array} \right) \nonumber \\
&\equiv& \hat{M}_{\widetilde{Q}_{\rm \normalsize L} ij}^2 + \Delta M_{\widetilde{Q}_{{\rm \normalsize L} ij}}^2, \label{massinsert} \\
M_{\widetilde{U}_{\rm \normalsize R} ij}^2 &\approx& \left( \begin{array}{ccc}
						a_2 \widetilde{M}^2 && \\
						& a_2 \widetilde{M}^2 & \\
						&& [a_2 + b_3 y_t^2] \widetilde{M}^2 \\
					\end{array} \right) \equiv \hat{M}_{\widetilde{U}_{\rm \normalsize R} i}^2, \\
M_{\widetilde{D}_{\rm \normalsize R} ij}^2 &\approx& \left( \begin{array}{ccc}
						[a_3 + b_4 y_d^2] \widetilde{M}^2 && \\
						& [a_3 + b_4 y_s^2] \widetilde{M}^2 & \\
						&& [a_3 + b_4 y_b^2] \widetilde{M}^2 \\
					\end{array} \right) \equiv \hat{M}_{\widetilde{D}_{\rm \normalsize R} i}^2, 
\end{eqnarray}

\begin{figure}[t]
\hspace{1.2cm}\includegraphics[width=.4\linewidth]{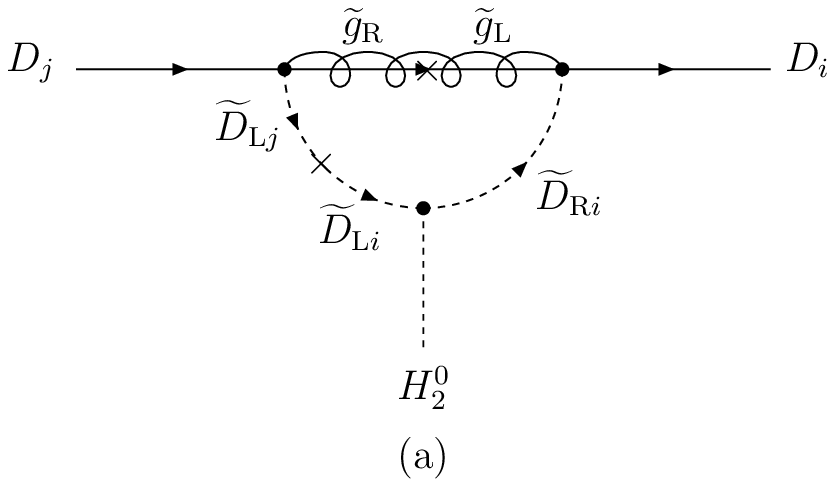}
\hspace{1.2cm}\includegraphics[width=.4\linewidth]{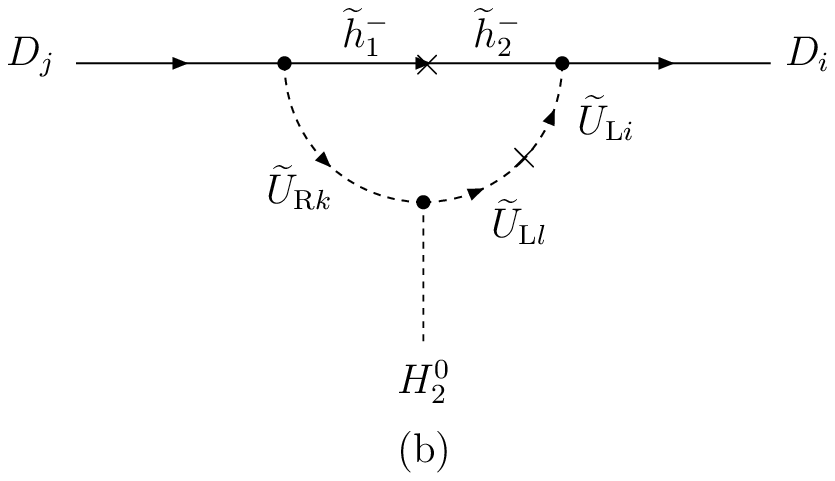}	
\caption{Subleading corrections to the down-type quark Yukawa couplings induced by (a) gluino $\widetilde{g}_{\normalsize \rm L,R}$ and (b) charged higgsino $\widetilde{h}^-_{1,2}$.}\label{fig:gluino_massin}
\end{figure} 

We calculate the correction to the down-type Yukawa coupling in Eq.(\ref{lagrangian}). In addition to the Fig.\ref{fig:gluino}, there are extra contributions shown in Fig.\ref{fig:gluino_massin}, where the off-diagonal terms in Eq.(\ref{massinsert}) are treated as mass insertion. The explicit form of $\Delta {\bf y}_d$ is given by
\begin{eqnarray}
\Delta {\bf y}_{d ij} &\approx& \left( \begin{array}{ccc} 
										\Delta {\bf y}_{d 11} & \Delta {\bf y}_{d 12} & \Delta {\bf y}_{d 13} \\
										\Delta {\bf y}_{d 21} & \Delta {\bf y}_{d 22} & \Delta {\bf y}_{d 23} \\
										\Delta {\bf y}_{d 31} & \Delta {\bf y}_{d 32} & \Delta {\bf y}_{d 33} \\
										\end{array} \right), \\
\Delta {\bf y}_{d 11} &\equiv& y_d \Eg^{(1)} \\
\Delta {\bf y}_{d 12} &\equiv& y_d [ \Egp^{(12)} + \Eh^{(13)} + \Ehp^{(133)} ]\KMtreedag_{13} \KMtree_{32}, \\
\Delta {\bf y}_{d 13} &\equiv& y_d [ \Egp^{(13)} + \Eh^{(13)} + \Ehp^{(133)} ] \KMtreedag_{13} \KMtree_{33}, \\
\Delta {\bf y}_{d 21} &\equiv& y_s [ \Egp^{(21)} + \Eh^{(23)} + \Ehp^{(233)} ] \KMtreedag_{23} \KMtree_{31}, \\
\Delta {\bf y}_{d 22} &\equiv& y_s \Eg^{(2)}, \\
\Delta {\bf y}_{d 23} &\equiv& y_s [ \Egp^{(23)} + \Eh^{(23)} + \Ehp^{(233)} ]\KMtreedag_{23} \KMtree_{33}, \\
\Delta {\bf y}_{d 31} &\equiv& y_b [ \Egp^{(31)} + \Eh^{(33)} ] \KMtreedag_{33} \KMtree_{31}, \\
\Delta {\bf y}_{d 32} &\equiv& y_b [ \Egp^{(32)} + \Eh^{(33)} ] \KMtreedag_{33} \KMtree_{32}, \\
\Delta {\bf y}_{d 33} &\equiv& y_b [ \Eg^{(3)} + \Eh^{(33)} ] \KMtreedag_{33} \KMtree_{33},
\end{eqnarray}
where,
\begin{eqnarray}
\Eg^{(i)} &\equiv& \frac{2 \alpha_s}{3 \pi}\frac{\mu^*}{M_{\widetilde{g}}}I^{(3)}\left[\frac{\hat{M}_{\widetilde{Q}_{\rm \normalsize L} i}}{M_{\widetilde{g}}}, \frac{\hat{M}_{\widetilde{D}_{\rm \normalsize R} i}}{M_{\widetilde{g}}}\right], \\
\Eh^{(i3)} &\equiv& -\frac{\mu}{16\pi^2}\frac{A_u}{M_{\widetilde{h}}} y_t^2 I^{(3)}\left[\frac{\hat{M}_{\widetilde{Q}_{\rm \normalsize L} i}}{M_{\widetilde{h}}}, \frac{\hat{M}_{\widetilde{U}_{\rm \normalsize R} 3}}{M_{\widetilde{h}}}\right], \\
\Egp^{(ij)} &\equiv& -\frac{2 \alpha_s}{3 \pi}\frac{\mu^*}{M_{\widetilde{g}}^3} b_1 y_t^2 \widetilde{M}^2 I^{(4)}\left[\frac{\hat{M}_{\widetilde{Q}_{\rm \normalsize L} i}}{M_{\widetilde{g}}}, \frac{\hat{M}_{\widetilde{Q}_{\rm \normalsize L} j}}{M_{\widetilde{g}}}, \frac{\hat{M}_{\widetilde{D}_{\rm \normalsize R} i}}{M_{\widetilde{g}}}\right], \\
\Ehp^{(il3)} &\equiv& \frac{\mu}{16\pi^2}\frac{A_u}{M_{\widetilde{h}}^3} b_1 y_t^4 \widetilde{M}^2 I^{(4)}\left[\frac{\hat{M}_{\widetilde{Q}_{\rm \normalsize L} i}}{M_{\widetilde{h}}}, \frac{\hat{M}_{\widetilde{Q}_{\rm \normalsize L} l}}{M_{\widetilde{h}}}, \frac{\hat{M}_{\widetilde{U}_{\rm \normalsize R} 3}}{M_{\widetilde{h}}}\right], \\
I^{(3)}[x,y] &=& \frac{x^2 \ln x^2}{(x^2 - 1)(x^2 - y^2)} + \frac{y^2 \ln y^2}{(y^2 - 1)(y^2 - x^2)} \\
I^{(4)}[x,y,z] &=& \frac{x^2 \ln x^2}{(x^2 - 1)(x^2 - y^2)(x^2 - z^2)} \nonumber \\
&&+ \frac{y^2 \ln y^2}{(y^2 - 1)(y^2 - x^2)(y^2 - z^2)} + \frac{z^2 \ln z^2}{(z^2 - 1)(z^2 - x^2)(z^2 - y^2)}
\end{eqnarray}
From the $\Delta {\bf y}_d$, we obtain the mass term for up-type and down-type quarks. 
\begin{eqnarray}
\mathcal{L}_{\rm \normalsize quark} = -\ODR_i [\hat{M}_{d ij} + \Delta_{ij}] \DL_j -\OUR_i M_{u ij} \UL_j + {\rm h.c.},
\end{eqnarray}
here
\begin{eqnarray}
\hat{M}_{d ij} &=& \left( \begin{array}{ccc}
					\hat{M}_{11} && \\
					& \hat{M}_{22} & \\
					&& \hat{M}_{33} \\
				\end{array} \right) \equiv \hat{M}_{d i}, \\
\hat{M}_{d 11} &\equiv&	\frac{v}{\sqrt{2}} \cos\beta y_d [1 + \Eg^{(1)}\tan\beta], \\
\hat{M}_{d 22} &\equiv& \frac{v}{\sqrt{2}} \cos\beta y_s [1 + \Eg^{(2)}\tan\beta], \\
\hat{M}_{d 33} &\equiv& \frac{v}{\sqrt{2}} \cos\beta y_b [1 + (\Eg^{(3)} + \Eh^{(33)})\tan\beta], \\
\Delta_{ij} &=& \left( \begin{array}{ccc}
						& \Delta_{12} & \Delta_{13} \\
						\Delta_{21} && \Delta_{23} \\
						\Delta_{31} & \Delta_{32} & \\
						\end{array} \right), \\
\Delta_{12} &\equiv& \frac{v}{\sqrt{2}}\cos\beta y_d [ \Egp^{(12)} + \Eh^{(13)} + \Ehp^{(133)} ] \KMtreedag_{13} \KMtree_{32} \tan\beta, \\
\Delta_{13} &\equiv& \frac{v}{\sqrt{2}}\cos\beta y_d [ \Egp^{(13)} + \Eh^{(13)} + \Ehp^{(133)} ] \KMtreedag_{13} \KMtree_{33}\tan\beta, \\
\Delta_{21} &\equiv& \frac{v}{\sqrt{2}}\cos\beta y_s [ \Egp^{(21)} + \Eh^{(23)} + \Ehp^{(233)} ] \KMtreedag_{23} \KMtree_{31} \tan\beta, \\
\Delta_{23} &\equiv& \frac{v}{\sqrt{2}}\cos\beta y_s [ \Egp^{(23)} + \Eh^{(23)} + \Ehp^{(233)} ] \KMtreedag_{23} \KMtree_{33} \tan\beta, \\
\Delta_{31} &\equiv& \frac{v}{\sqrt{2}}\cos\beta y_b [ \Egp^{(31)} + \Eh^{(33)} ] \KMtreedag_{33} \KMtree_{31} \tan\beta, \\
\Delta_{32} &\equiv& \frac{v}{\sqrt{2}}\cos\beta y_b [ \Egp^{(32)} + \Eh^{(33)} ] \KMtreedag_{33} \KMtree_{32} \tan\beta, \\
\hat{M}_u &\equiv& \frac{v}{\sqrt{2}}\sin\beta \hat{\bf y}_u \KMtree.
\end{eqnarray}
Next, we rotate the basis to mass eigen-states as follows:
\begin{eqnarray}
\UL &=& V_{\rm \normalsize L}(U) \UL^\prime = \KMtreedag \UL^\prime,\ \ \  \UR = V_{\rm \normalsize R}(U) \UR^\prime = \UR^\prime, \\
\DL &=& V_{\rm \normalsize L}(D) \DL^\prime,\ \ \  \DR = V_{\rm \normalsize R}(D) \DR^\prime.
\end{eqnarray}
We introduce $\Delta V_{\rm \normalsize L}$ and $\Delta V_{\rm \normalsize R}$ as
\begin{eqnarray}
V_{\rm \normalsize L}(D) = {\bf 1} + \Delta V_{\rm \normalsize L}, \\
V_{\rm \normalsize R}(D) = {\bf 1} + \Delta V_{\rm \normalsize R},
\end{eqnarray}
where the unitarity requires that $\Delta V_{\rm \normalsize L(R)}^\dag = -\Delta V_{\rm \normalsize L(R)}$ and $\Delta V_{{\rm \normalsize L(R)}ii} = 0$.
At the first order of $\Delta_{ij}$, $\Delta V_{\rm \normalsize L,R}$ is expressed as \cite{Isidori:2001fv}

\begin{eqnarray}
\Delta V_{{\rm \normalsize L} ij} &=& - \frac{\hat{M}_{d i} \Delta_{ij} + \Delta_{ij}^\dag \hat{M}_{d j}}{\hat{M}_{d i}^2 - \hat{M}_{d j}^2}\ \ \ \textrm{for $i \neq j$}, \\
\Delta V_{{\rm \normalsize R} ij} &=& - \frac{\hat{M}_{d i} \Delta_{ij}^\dag + \Delta_{ij} \hat{M}_{d j}}{\hat{M}_{d i}^2 - \hat{M}_{d j}^2}\ \ \ \textrm{for $i \neq j$}. \label{delta_V_R}
\end{eqnarray}
At this order $\hat{M}_d$ is the physical down-type quark mass matrix, since there are no corrections to diagonal terms.

We derive relationship between $\KMtree$ and $\KM$. Since the $W$ boson coupling is given by
\begin{eqnarray}
\mathcal{L}_{W^\pm} &=& \frac{g_2}{\sqrt{2}}[ \overline{U}_{\rm \normalsize L} W^+_\mu \gamma^\mu D_{\rm \normalsize L} + {\rm h.c.} ] \nonumber \\
&=& \frac{g_2}{\sqrt{2}}[ \overline{U}_{\rm \normalsize L}^\prime W^+_\mu \gamma^\mu \KMtree V_{\rm \normalsize L}(D) D_{\rm \normalsize L}^\prime + {\rm h.c.} ] \nonumber \\
&=& \frac{g_2}{\sqrt{2}}[ \overline{U}_{\rm \normalsize L}^\prime W^+_\mu \gamma^\mu \KM D_{\rm \normalsize L}^\prime + {\rm h.c.} ],
\end{eqnarray}
we obtain
\begin{eqnarray}
\KM \equiv \KMtree V_{\rm \normalsize L}(D).
\end{eqnarray}
Explicit form for each elements are given by
\begin{eqnarray}
\KMtree_{11} &=& \KM_{11}, \\ \label{VKM_VKM_1}
\KMtree_{12} &=& \KM_{12}, \\
\KMtree_{13} &=& \KM_{13}\frac{1 + [\Eg^{(3)} + \Eh^{(33)}]\tan\beta}{1 + [ \Eg^{(3)} + \Eh^{(33)} + \Egp^{(13)} + \Eh^{(13)} + \Ehp^{(133)} ]\tan\beta}, \\
\KMtree_{21} &=& \KM_{21}, \\
\KMtree_{22} &=& \KM_{22}, \\
\KMtree_{23} &=& \KM_{23}\frac{1 + [\Eg^{(3)} + \Eh^{(33)}]\tan\beta}{1 + [ \Eg^{(3)} + \Eh^{(33)} + \Egp^{(23)} + \Eh^{(23)} + \Ehp^{(233)} ]\tan\beta}, \\
\KMtree_{31} &=& \KM_{31} \frac{1 + [\Eg^{(3)} + \Eh^{(33)}]\tan\beta}{1 + [\Eg^{(3)} - \Egp^{31}]\tan\beta}, \\
\KMtree_{32} &=& \KM_{32} \frac{1 + [\Eg^{(3)} + \Eh^{(33)}]\tan\beta}{1 + [\Eg^{(3)} - \Egp^{(32)}]\tan\beta}, \\
\KMtree_{33} &=& \KM_{33}. \label{VKM_VKM_2}
\end{eqnarray}

Then the charged Higgs couplings can be expressed as
\begin{eqnarray}
\mathcal{L}_{H^\pm} &=& \sin\beta H^-\ODR_i \hat{\bf y}_{d ij}\UL_j + {\rm h.c.} \nonumber \\
&=& \sin\beta H^- \ODR^\prime_i V_{\rm \normalsize R}^\dag(D)_{ik} \hat{\bf y}_{d k} V_{\rm \normalsize L}(U)_{kj} \UL_j^\prime + {\rm h.c.} \nonumber \\
&=& \sin\beta H^- \ODR^\prime_i V_{\rm \normalsize R}^\dag(D)_{ik} \hat{\bf y}_{d k} \KMtreedag_{kj} \UL_j^\prime + {\rm h.c.}.
\end{eqnarray}
Using Eqs.(\ref{delta_V_R}) and (\ref{VKM_VKM_1}) - (\ref{VKM_VKM_2}), we obtain\begin{eqnarray}
\mathcal{L}_{H^\pm} &\approx& \frac{\sqrt{2}}{v}\tan\beta H^- \ODR^\prime_i \frac{\hat{M}_{d i}}{ 1 + [\Eg^{(i)} ] \tan\beta} \KMdag_{ij} \UL_j^\prime + {\rm h.c.} \ \ \ \ \ \ \textrm{for $(i,j)=(1,1),(1,2),(2,1),(2,2)$}, \nonumber \\
\\
\mathcal{L}_{H^\pm} &\approx& \frac{\sqrt{2}}{v}\tan\beta H^- \ODR^\prime_i \frac{\hat{M}_{d i}}{ 1 + [\Eg^{(i)} - \Egp^{(ij)} ] \tan\beta} \KMdag_{ij} \UL_j^\prime + {\rm h.c.}  \ \ \ \ \ \ \textrm{for $(i,j)=(3,1),(3,2)$}, \nonumber \\
\\
\mathcal{L}_{H^\pm} &\approx& \frac{\sqrt{2}}{v}\tan\beta H^- \ODR^\prime_i \frac{\hat{M}_{d i}}{1 + \Eg^{(i)}\tan\beta } \frac{ 1 + [ \Eg^{(3)} + \Eh^{(33)} ]\tan\beta }{1 + [ \Eg^{(i)} + \Eh^{(33)} + \Egp^{(ij)} + \Eh^{(i3)} + \Ehp^{(i33)} ]\tan\beta} \KMdag_{ij} \UL_j^\prime \nonumber \\
&&  \ \ \ \ \ \ \ \ \ \ \ \ \ \ \ \ \ \ \ \ \ \ \ \ \ \ \ \ \ \ \ \ \ \ \ \ \ \ \ \ \ \ \ \ \ \ \ \ \ \ \ \ \ \ \ \ \ \ \ \ \ \ \ \   + {\rm h.c.}\ \ \ \ \ \  \textrm{for $(i,j)=(1,3),(2,3)$}, \\
\mathcal{L}_{H^\pm} &\approx& \frac{\sqrt{2}}{v}\tan\beta H^- \ODR^\prime_i \frac{\hat{M}_{d i}}{ 1 + [\Eg^{(i)} + \Eh^{(i3)} ] \tan\beta} \KMdag_{ij} \UL_j^\prime + {\rm h.c.}\ \ \ \ \ \ \textrm{for $(i,j)=(3,3)$}.
\end{eqnarray}
Notice that $b \to c$ and $b \to u$ transition do not receive the higgsino loop contribution just as in the simplest case discussed in section 2.

%

\end{document}